\documentclass[doublespacing]{elsart}
\usepackage{graphicx}
\usepackage{amsmath}
\usepackage{amsfonts}
\usepackage{amssymb}

\begin{document}

\begin{frontmatter}
\title{How to count trees?}
\author{Sebastian Piec,}
\author{Krzysztof Malarz\corauthref{km}}
\ead{malarz@agh.edu.pl}
\ead[url]{http://home.agh.edu.pl/malarz/}
\corauth[km]{Corresponding author. Fax: +48 12 6340010.}
\and
\author{Krzysztof Ku{\l}akowski}
\address{
AGH University of Science and Technology,
Faculty of Physics and Applied Computer Science,
al. Mickiewicza 30, PL-30059 Krak\'ow, Poland}

\begin{abstract}
We propose a new topological invariant of unlabeled trees of $N$ nodes.
The invariant is a set of $N\times 2$ matrices of integers, with $\sum_j k^{d_{i,j}}$ 
and $v_i$ as the matrix elements, where $d_{i,j}$ are the elements of the distance 
matrix and $v_i$ denotes $i$-th node's degree and $k\in\mathbb{N}$.
To compare the invariant calculated for possibly different graphs, the matrix rows 
are ordered with respect to first column, and --- if necessary --- with respect to 
the second one.
We use the new invariant to evaluate from below the number of topologically different 
unlabeled trees up to $N=17$.
The results slightly exceed the asymptotic evaluation of Otter. 
\end{abstract}


\end{frontmatter}

\section{Introduction}
Averaging over different graphs is basic in numerous applications of the graph theory 
\cite{graph,tab8}.
For such tasks, knowledge of the number of topologically different graphs is of primary 
importance.
Having two graphs, a typical question is: are they different?
If the graphs are labeled, respective algorithms are of polynomial time.
However, for unlabeled graphs the task should be to check all possible labellings, what 
makes the problem unfeasible \cite{free}.
An alternative solution is to find a quantity which is different for different graphs, 
and of the same value if the graphs are topologically equivalent.
The latter means that there is a one-to-one transformation from one graph to another: 
each pair of nodes linked (not linked) in one graph is linked (not linked) in another graph.
Such a quantity is a topological invariant. 
However, actually we can be never sure if the quantity proposed as the invariant has indeed 
the above discriminating property.
While its different values certainly mean different graph structures, the same value does 
not allow to claim that the graphs are indeed topologically identical.
In many cases, the proposed quantity appears to be degenerate, i.e. its value is the same 
for different graphs.
All that remains true for unlabeled trees, which are graphs without cyclic paths and without 
loops.

In a series of papers, Schultz et al. proposed and evaluated some scalar quantities
as candidates to be topological invariants for trees \cite{szulc}. This work was 
motivated by a chemical application of the constructed quantities, which were found to 
increase monotonically with the melting temperature of alkanes. However, almost all 
proposed invariants were found to be degenerate. On the other hand, the last proposed 
invariant is a real number and not integer, and the comparison of its value must rely on
the numerical accuracy.

Here we propose a new candidate as a topological invariant for unlabeled trees. Unlike the
quantities discussed previously, this is a set of matrices and not a single number. The advantage
is that the matrices are ordered in a simple way, and the ordering algorithm works in 
polynomial time. On the other hand, to state that two trees are topologically identical 
we compare all the matrix elements. This modification is expected to enhance the discriminative 
force of the proposed invariant. We use the obtained criterion to calculate the number 
of topologically non-equivalent trees up to $N=17$ nodes. As stated above, the obtained
numbers can be treated only as an evaluation of the true results from below. Then,
if one has a better criterion, he should find the greater number of trees for $N\le 17$, 
than our result, given in Table 1. 

\begin{table}
\caption{\label{tab} The number of trees $T$ evaluated basing on sorted $({\bf b},{\bf v})$ 
pairs with $k\le 6$.
$T_\text{O}$ is given by the Otter's formula \eqref{eq-otter}.}
\begin{center}
\begin{tabular}{l rrrrr rrrrr}
\hline \hline
$N$ &
           1 &
           2 &
           3 &
           4 &
           5 \\
$T$ &
           1 &
           1 &
           1 &
           2 &
           3 \\
$T_{\text{O}}$ &
         1.6 &
         0.8 &
         0.9 &
         1.3 &
         2.2 \\
\hline
             &
           6 &
           7 &
           8 &
           9 &
          10 \\
             &
           6 &
          11 &
          23 &
          47 &
         106 \\
             &
         4.0 &
         8.1 &
        17.2 &
        37.9 &
        86.1 \\
\hline
             &
          11 &
          12 &
          13 &
          14 &
          15 \\
             &
         235 &
         551 &
        1301 &
        3159 &
        7741 \\
             &
       200.5 &
       476.9 &
      1153.9 &
      2833.8 &
      7049.1 \\
\hline
             &
          16 &
          17 \\ 
             &
       19320 &
       48629 \\ 
             &
     17731.0 &
     45038.0 \\  
\hline \hline
\end{tabular}
\end{center}
\end{table}

In our next section, our numerical procedure is described in details. Section \ref{sec-3} contains the 
numerical results. The obtained numbers of trees are compared to the analytical
evaluation of Otter \cite{otter}. In Section \ref{sec-4} we provide an argument that the range of 
values of any good candidate of a topological invariant should increase exponentially 
with the number of nodes $N$. Our proposition is the only one we know to fulfill this criterion. 
However, this `criterion of range' is not sufficient in the sense that it does not exclude the 
possible degeneracy. 

\section{Numerical approach}

Our numerical approach is based on the construction of the distance matrix $\mathbf{D}^N$ during tree 
growth \cite{km}.
In distance matrix ${\bf D}$ element $d_{i,j}$ gives the length of the shortest path between 
nodes $i$ and $j$, i.e. the minimal number of edges which connect these vertices.
The construction algorithm relies on the fact that a distance to a newly added $(N+1)$-th node 
to all other nodes $1\le i\le N$ via node $q$ --- to which new node is attached --- is 
$d_{N+1,i}=d_{q,i}+1$.
The computational complexity of the distance matrix ${\bf D}$ construction recipe is of order 
of $O(N^2)$.
The number of `1' in $i$-th row gives $i$-th node's degree $v_i$.

For counting trees two single-column vectors seem to be useful:
the first one $\mathbf{b}$ gives sum of the natural parameter
$k\in\mathbb{N}$ to the power equal to distance $d_{i,j}$ of $i$-th node to
another node $j$: $b_i=\sum_{j=1}^N k^{d_{i,j}}$.
The second vector ${\bf v}$ serves node's degrees $v_i$.
These vectors form a matrix, which is sorted with key pair
$(\mathbf{b},\mathbf{v})$: two trees are different if their
$(\mathbf{b},\mathbf{v})$ are different for {\em all} values of $k$.
Actually, we compare the matrices for $k=2, 3, 4, 5$ and $6$.
We have checked numerically, that the results of trees counting are
different for $k\le 4$ and $k=5$ but they are the same for $k=5$ and
$k=6$.
Sorting elements of $(\mathbf{b},\mathbf{v})$ makes the matrix independent
on an order of labeling of the tree's nodes.

\begin{figure}
\begin{center}
(a) \includegraphics[scale=.4]{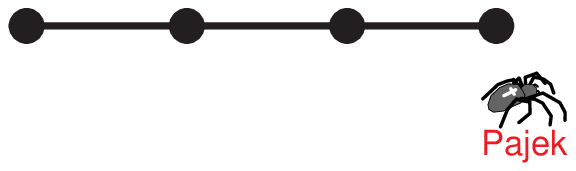}
(b) \includegraphics[scale=.4]{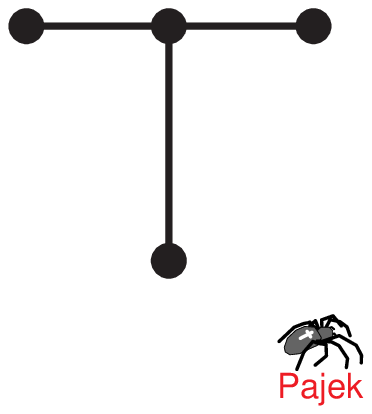}
\end{center}
\caption{\label{fig-tree4} Two trees of $N=4$ nodes.
(Figures using Pajek \cite{pajek}.)}
\end{figure}

\begin{figure}
\begin{center}
\includegraphics[scale=.3]{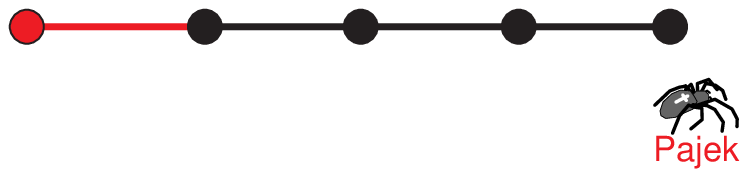}
\includegraphics[scale=.3]{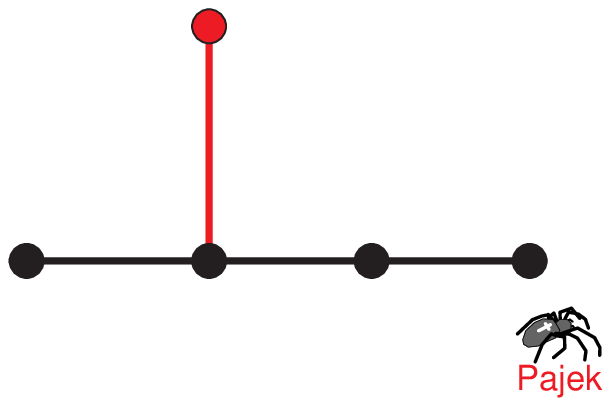}
\includegraphics[scale=.3]{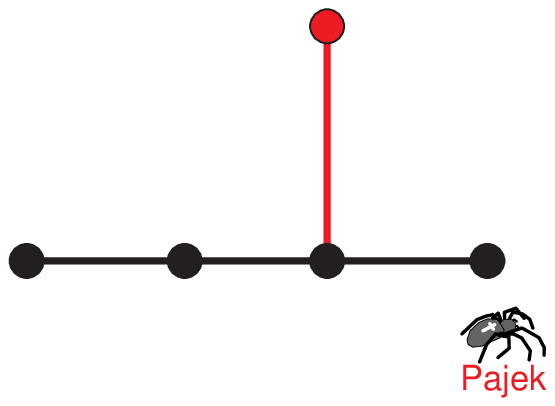}
\includegraphics[scale=.3]{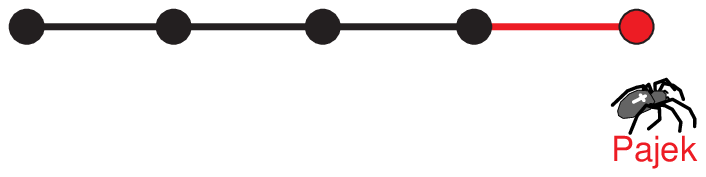}\\
\includegraphics[scale=.3]{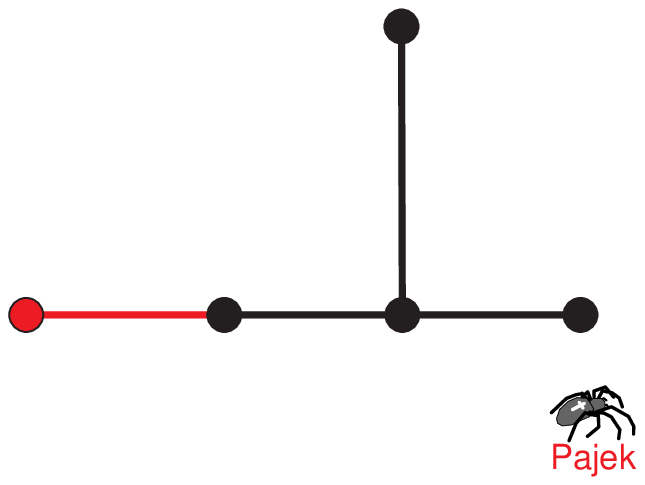}
\includegraphics[scale=.3]{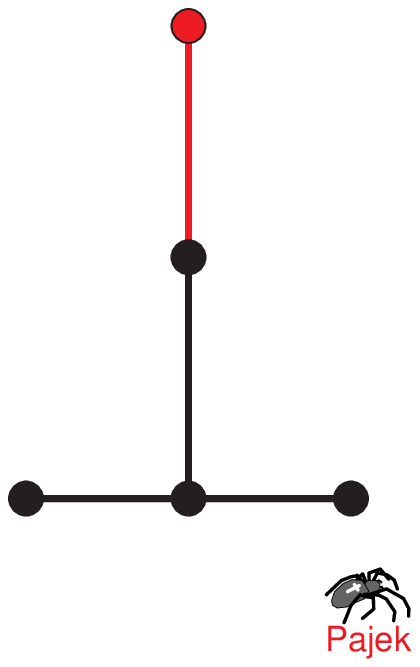}
\includegraphics[scale=.3]{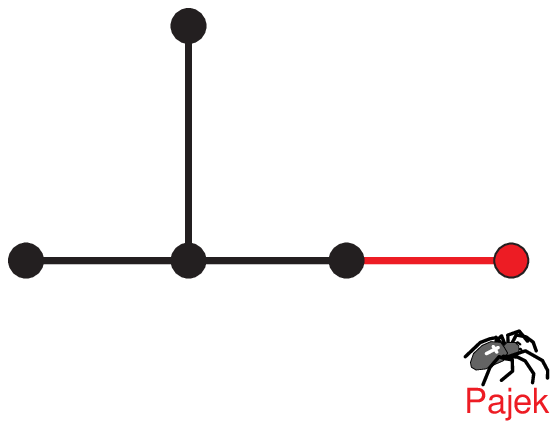}
\includegraphics[scale=.3]{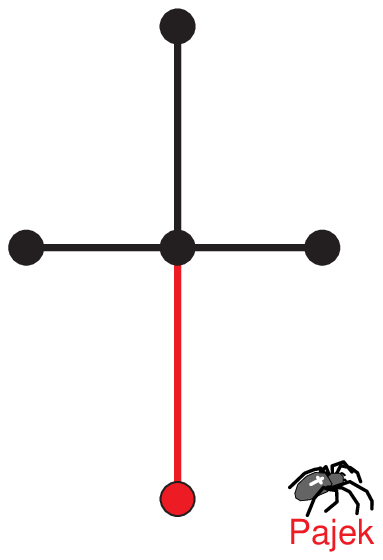}\\
(a) \includegraphics[scale=.3]{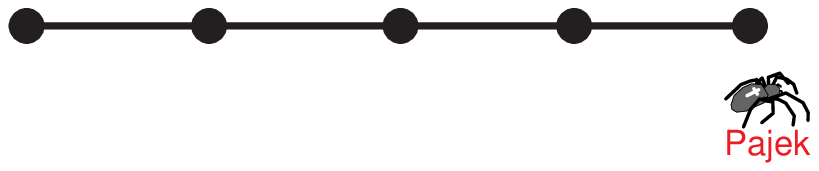}
(b) \includegraphics[scale=.3]{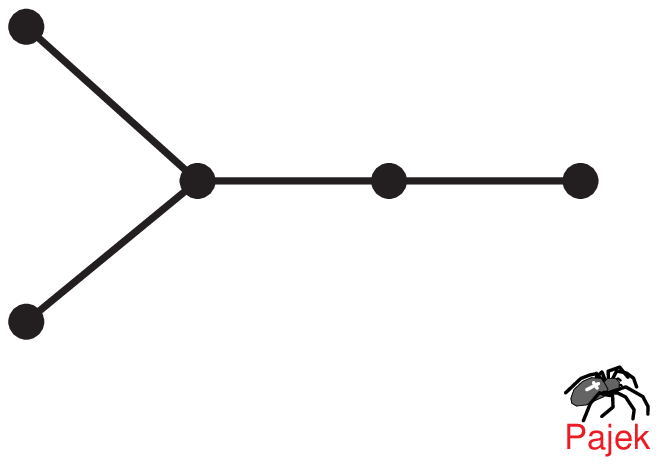}
(c) \includegraphics[scale=.3]{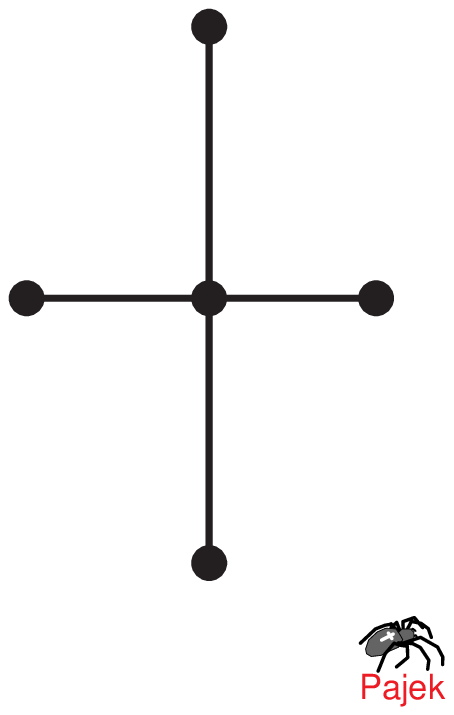}
\end{center}
\caption{\label{fig-tree5} Eight possible trees obtained by adding one node by one link in 
all possible ways to two trees with $N=4$.
Among them {\em only three} are different.}
\end{figure}

For example, the only two existing trees for $N=4$ --- presented in Fig. \ref{fig-tree4} --- 
have distance matrices $\mathbf{D}^4$ \cite{physicaa}:
\[ 
\mathbf{D}^4_a=
\begin{pmatrix}
0&1&2&3\\
1&0&1&2\\
2&1&0&1\\
3&2&1&0\\
\end{pmatrix}
\text{ and }
\mathbf{D}^4_b=
\begin{pmatrix}
0&1&2&2\\
1&0&1&1\\
2&1&0&2\\
2&1&2&0\\
\end{pmatrix}
\]
and sorted pair $(\mathbf{b},\mathbf{v})^4$ for $k=2$:
\[
(\mathbf{b},\mathbf{v})^4_a=
\begin{pmatrix}
15 & 1\\
15 & 1\\
 9 & 2\\
 9 & 2\\
\end{pmatrix}
\text{ and }
(\mathbf{b},\mathbf{v})^4_b=
\begin{pmatrix}
11 & 1\\
11 & 1\\
11 & 1\\
 7 & 3\\
\end{pmatrix}.
\]
Now, the next generation of trees is produced $N\to N+1$ by systematically adding a new node to 
each node for all preexisting trees.
For example in case of $N=4\to N=5$ look at Fig. \ref{fig-tree5}.
Among eight cases only three classes of $(\mathbf{b},\mathbf{v})^5$ exist, i.e.:
\[
(\mathbf{b},\mathbf{v})^5_a=
\begin{pmatrix}
31 & 1\\
31 & 1\\
17 & 2\\
17 & 2\\
13 & 2\\
\end{pmatrix},
(\mathbf{b},\mathbf{v})^5_b=
\begin{pmatrix}
23 & 1\\
19 & 1\\
19 & 1\\
13 & 2\\
11 & 3\\
\end{pmatrix}
\text{ and }
(\mathbf{b},\mathbf{v})^5_c=
\begin{pmatrix}
15 & 1\\
15 & 1\\
15 & 1\\
15 & 1\\
 9 & 4\\
\end{pmatrix}
\]
for $k=2$.
Three distance matrices $\mathbf{D}^5_{a,b,c}$ for these three trees are necessary to next step, 
i.e. $N=5\to N=6$.
The procedure is repeated recursively.

Technically, the sorting with key procedure is an implementation of the quick-sort algorithm 
\cite{quicksort} while comparing two $({\bf b},{\bf v})$ matrices are realized with standard 
C++ STL library \cite{stl}.

\section{Results of simulations}
\label{sec-3}
\begin{figure*}
\begin{center}
\includegraphics[scale=.25]{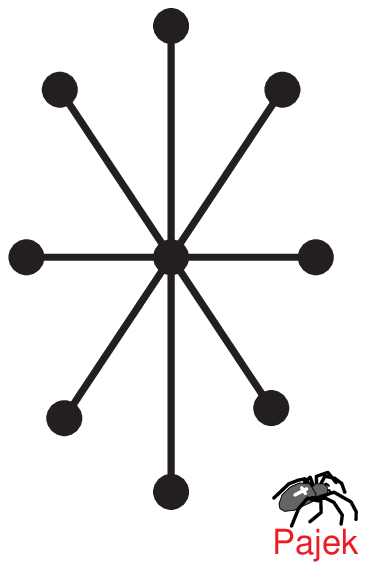}
\includegraphics[scale=.25]{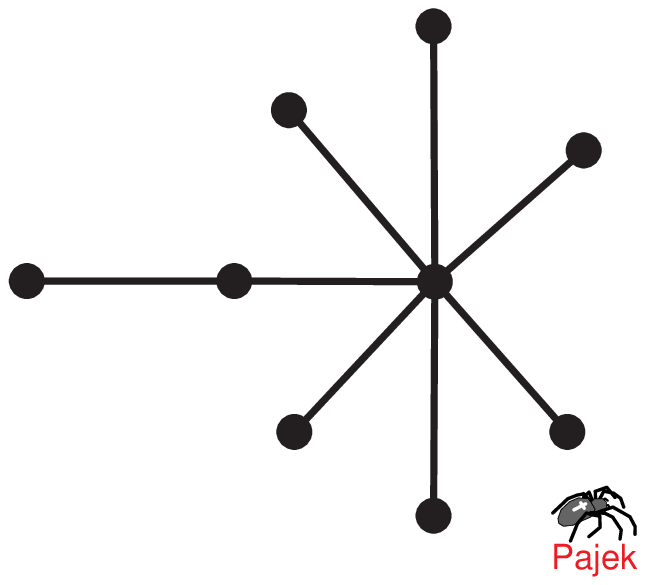}
\includegraphics[scale=.25]{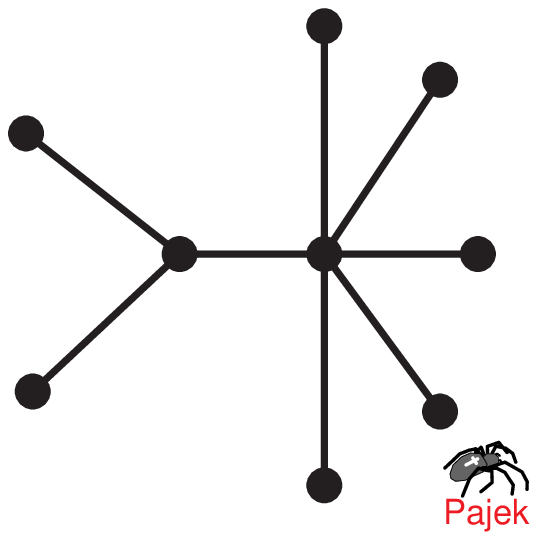}
\includegraphics[scale=.25]{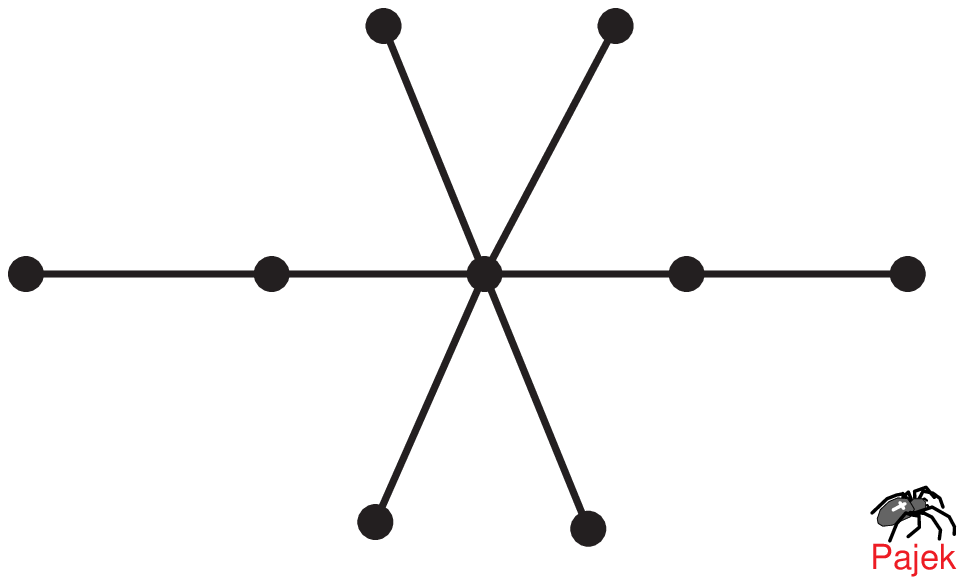}
\includegraphics[scale=.25]{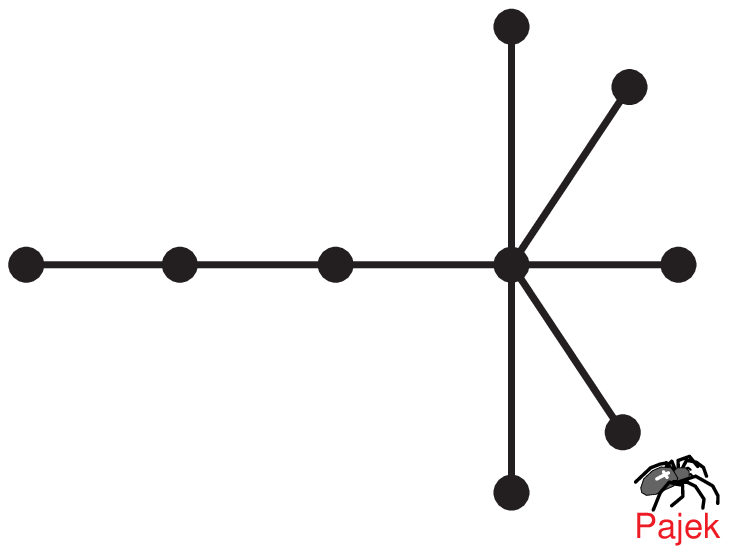}
\includegraphics[scale=.25]{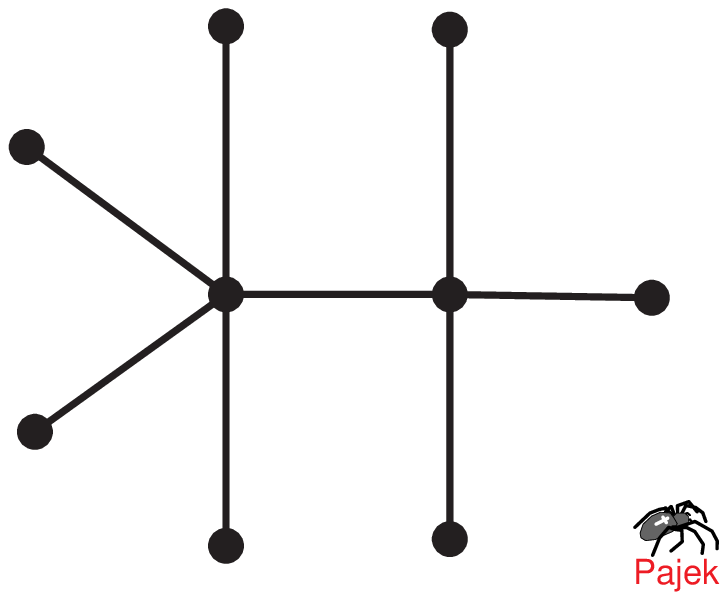}
\includegraphics[scale=.25]{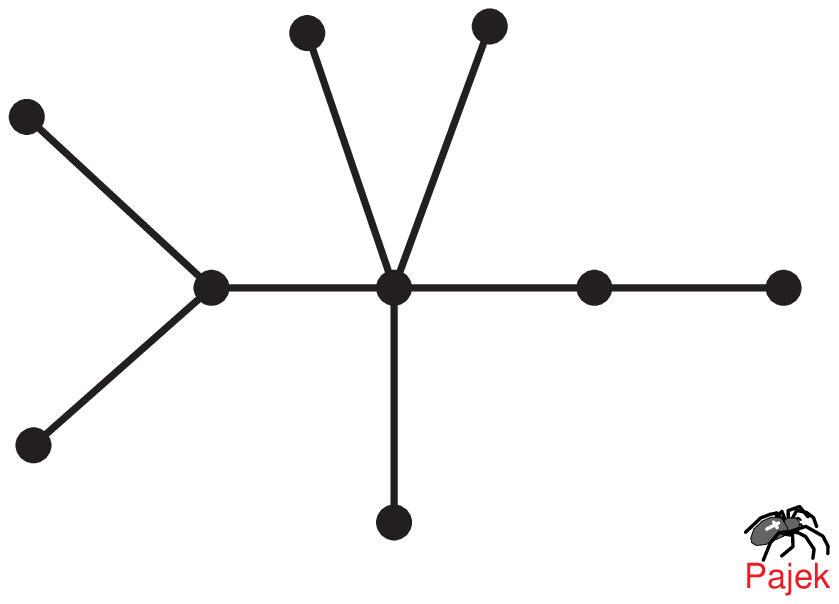}
\includegraphics[scale=.25]{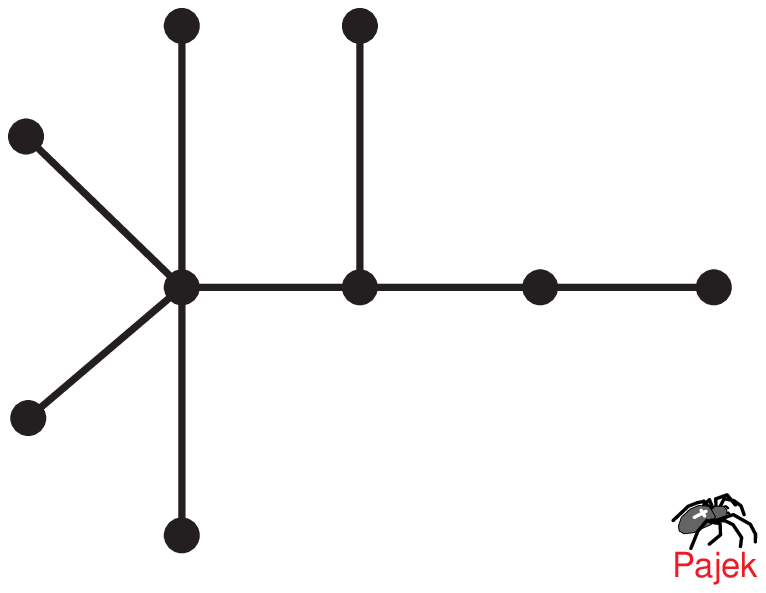}
\includegraphics[scale=.25]{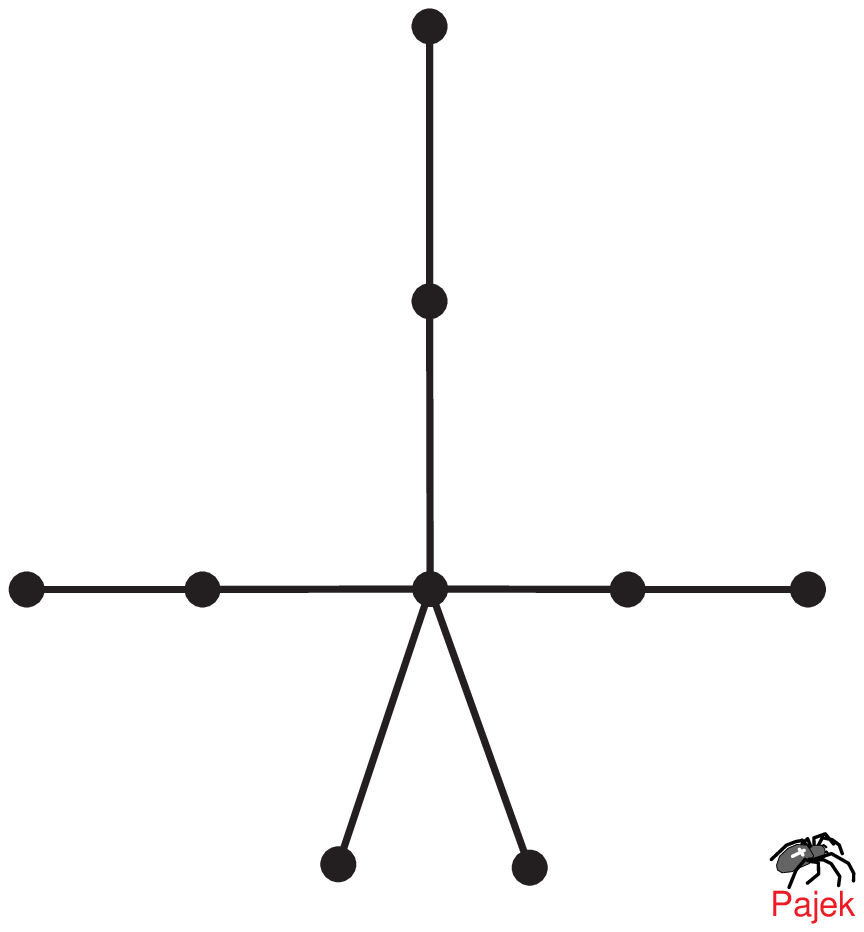}
\includegraphics[scale=.25]{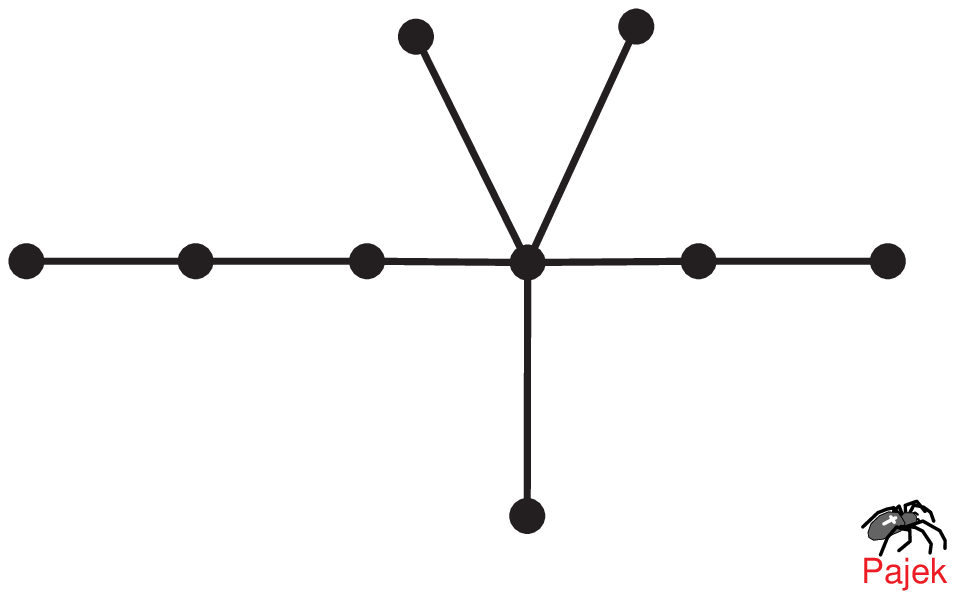}
\includegraphics[scale=.25]{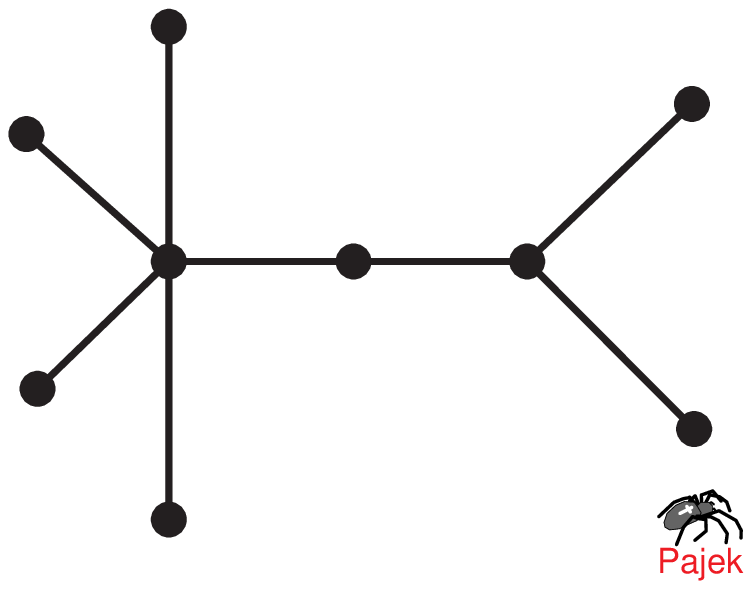}
\includegraphics[scale=.25]{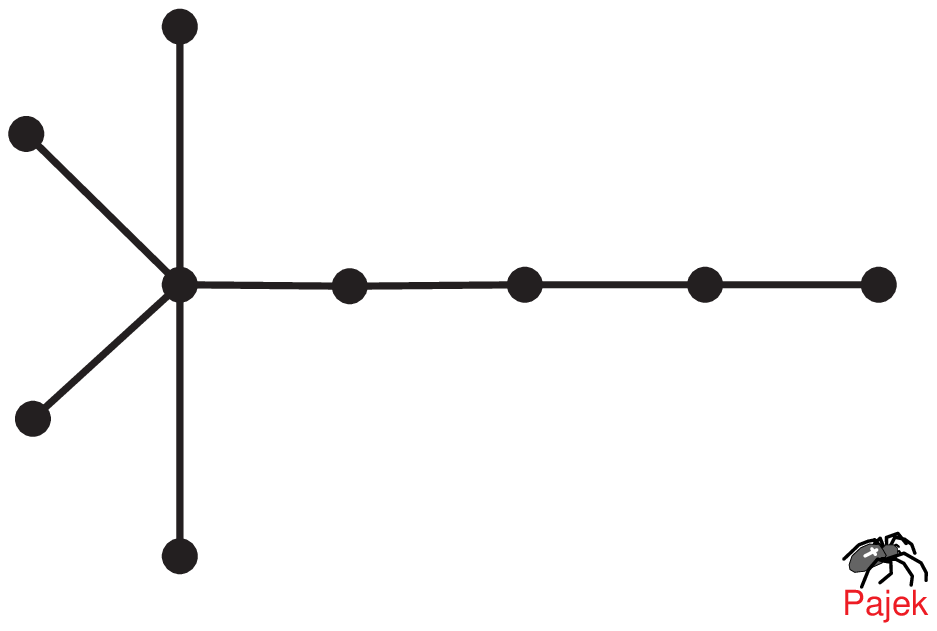}
\includegraphics[scale=.25]{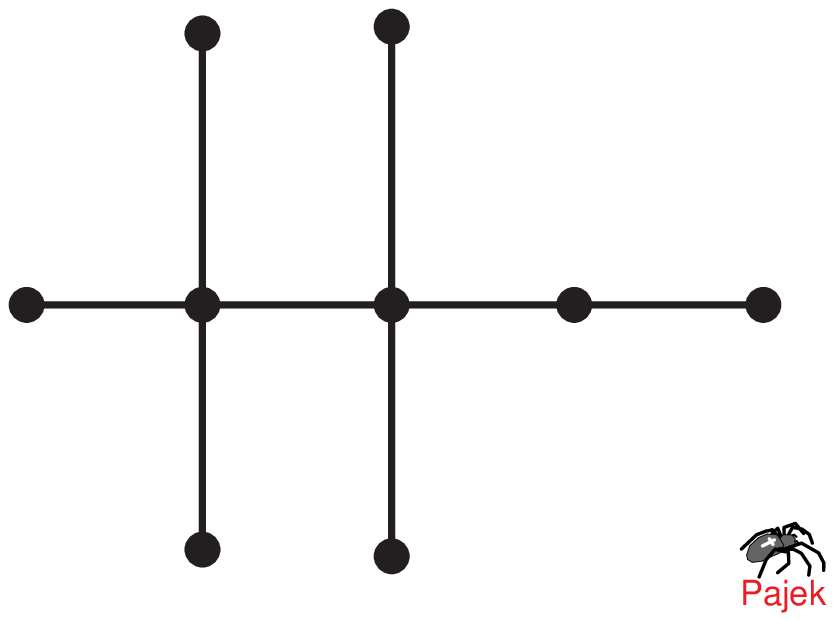}
\includegraphics[scale=.25]{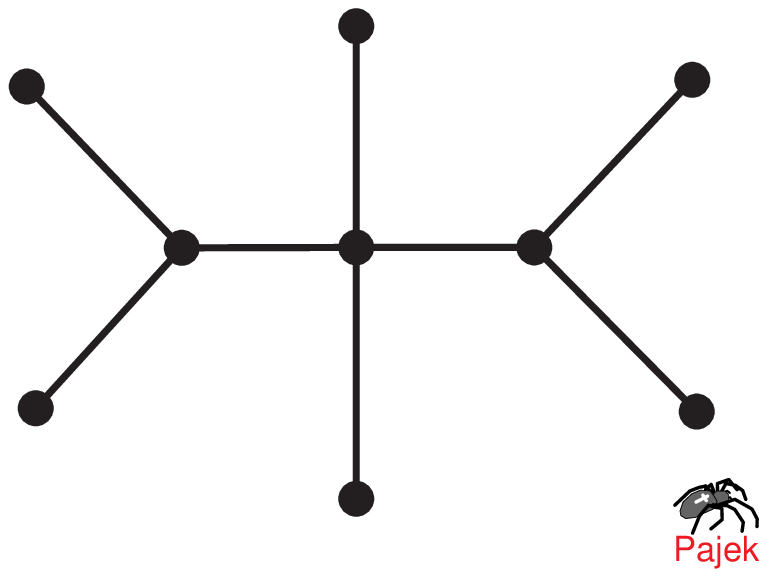}
\includegraphics[scale=.25]{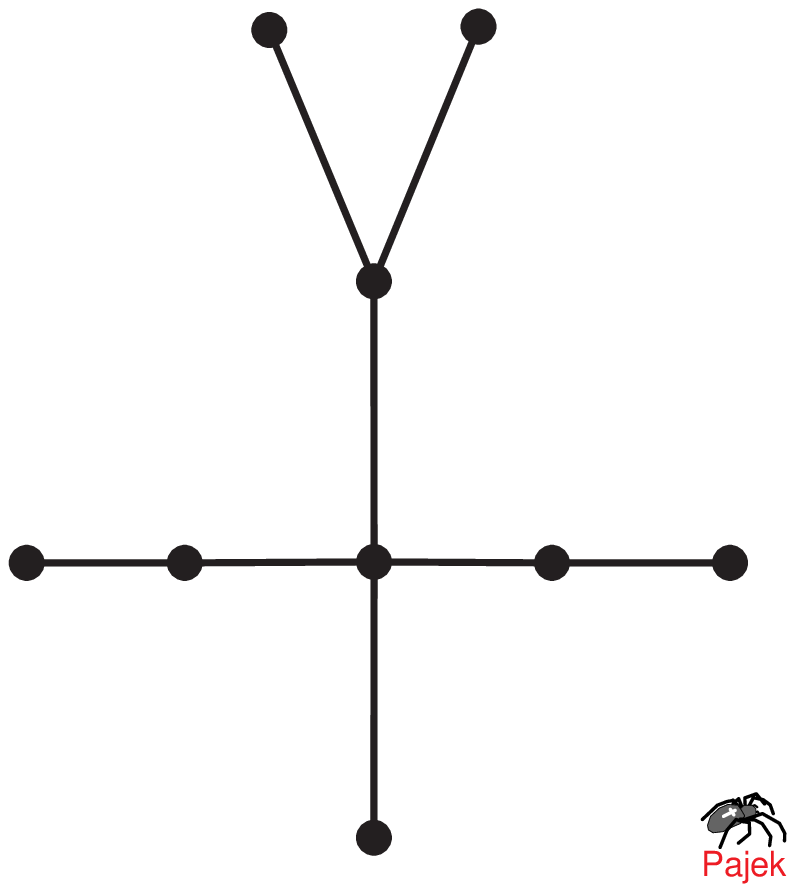}
\includegraphics[scale=.25]{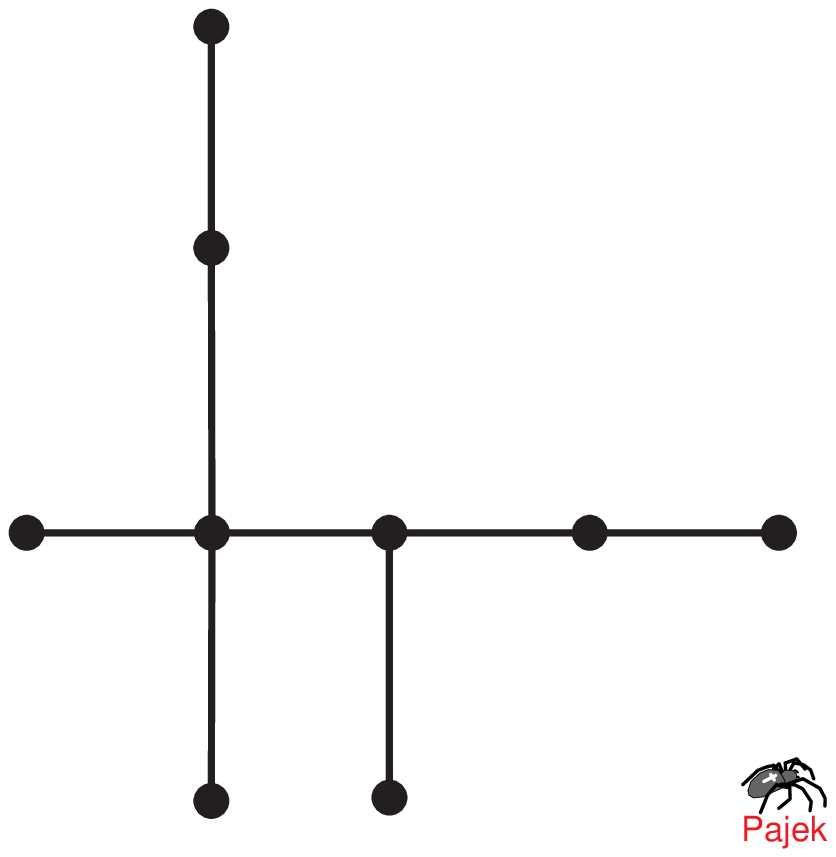}
\includegraphics[scale=.25]{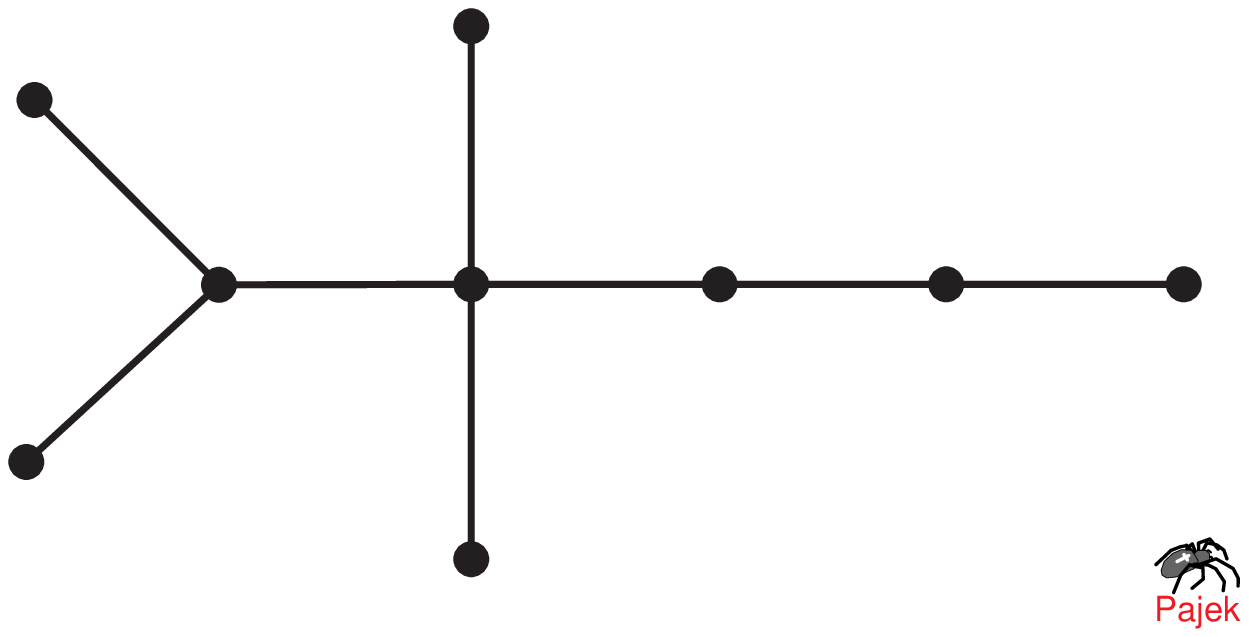}
\includegraphics[scale=.25]{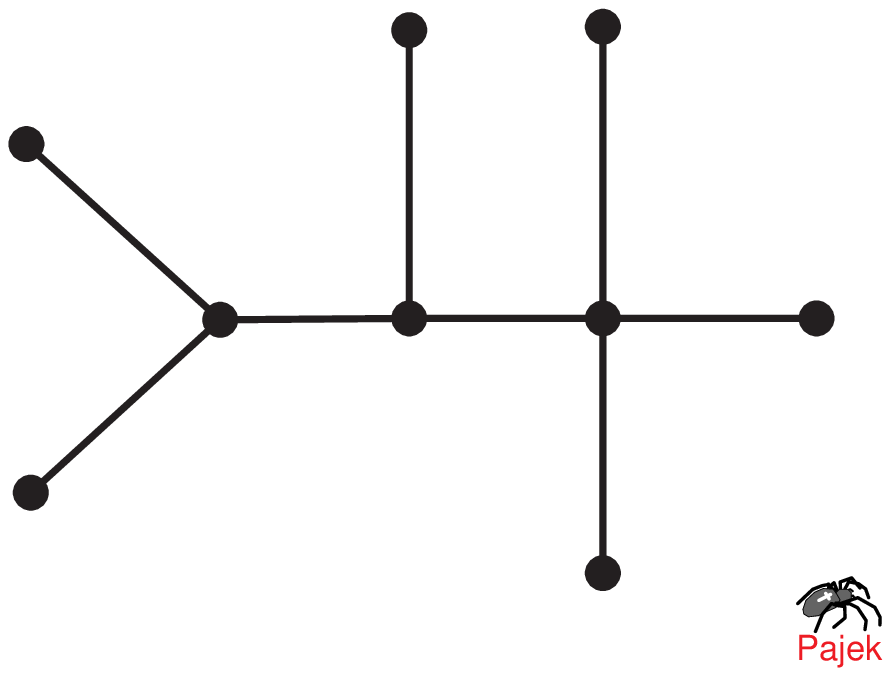}
\includegraphics[scale=.25]{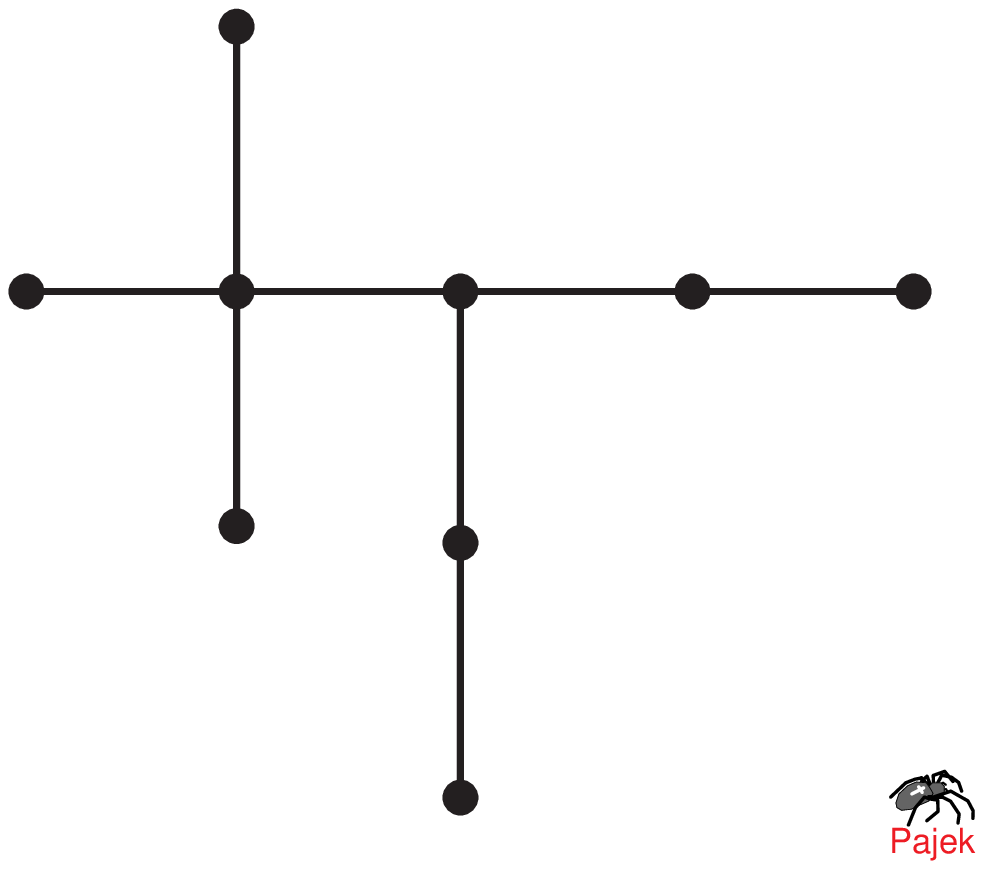}
\includegraphics[scale=.25]{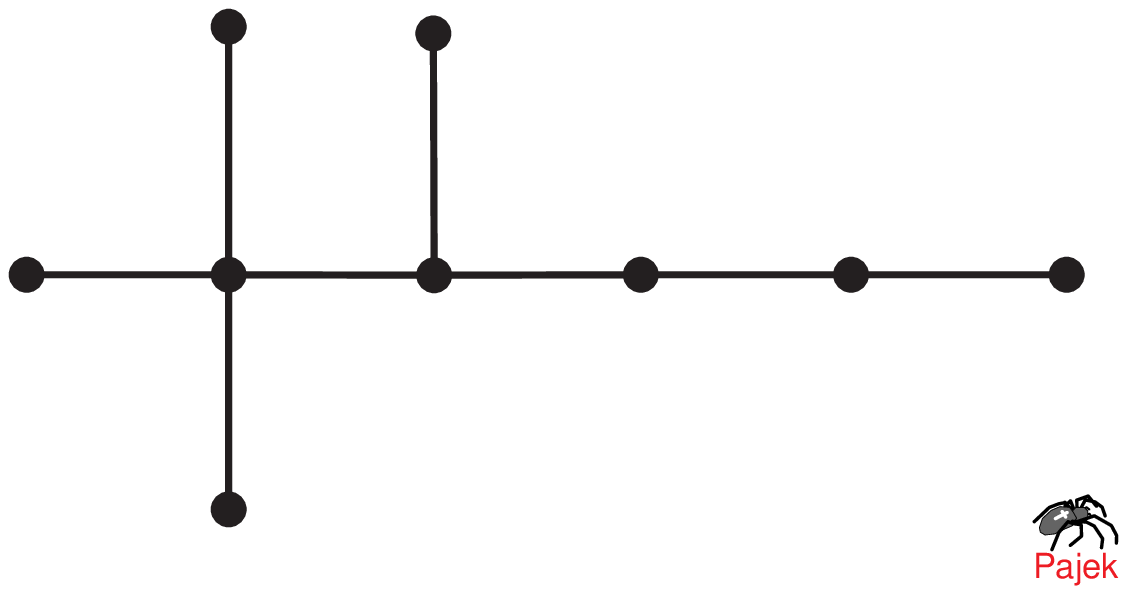}
\includegraphics[scale=.25]{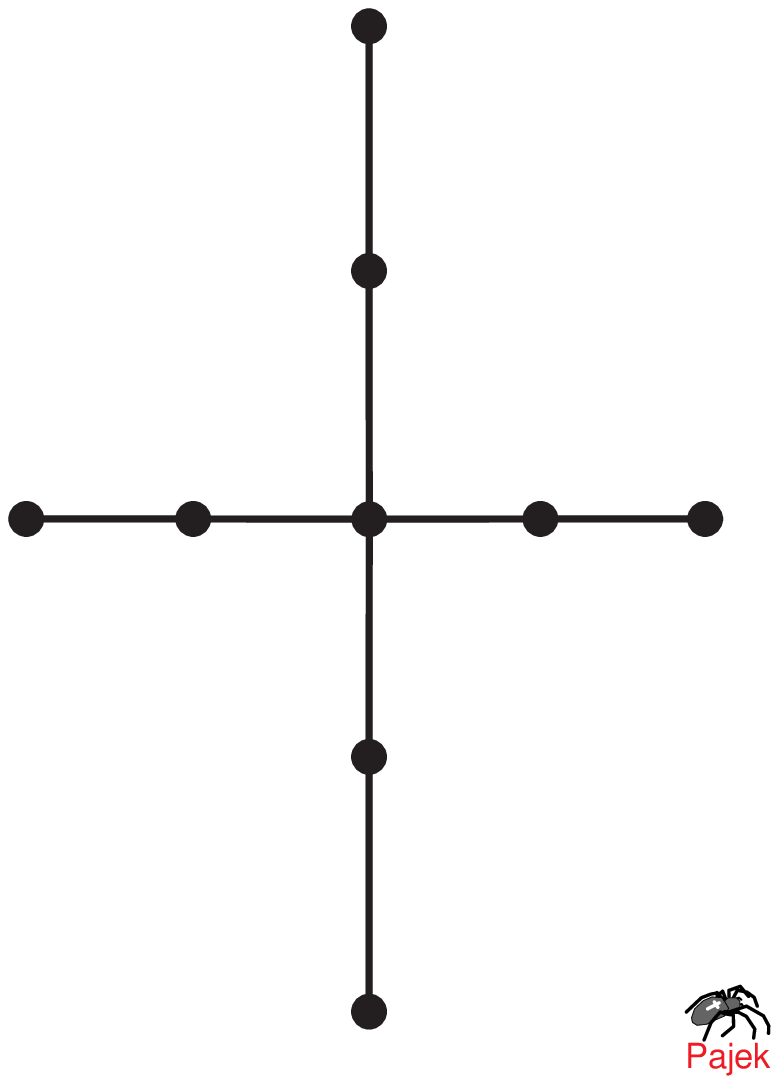}
\includegraphics[scale=.25]{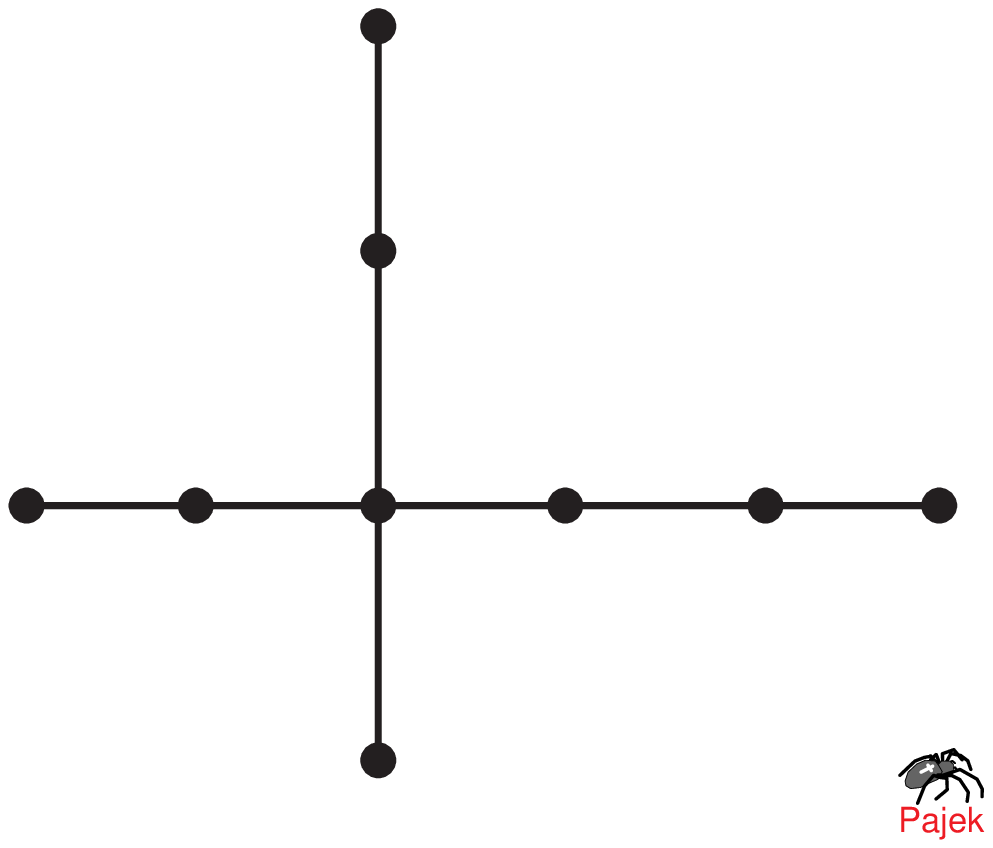}
\includegraphics[scale=.25]{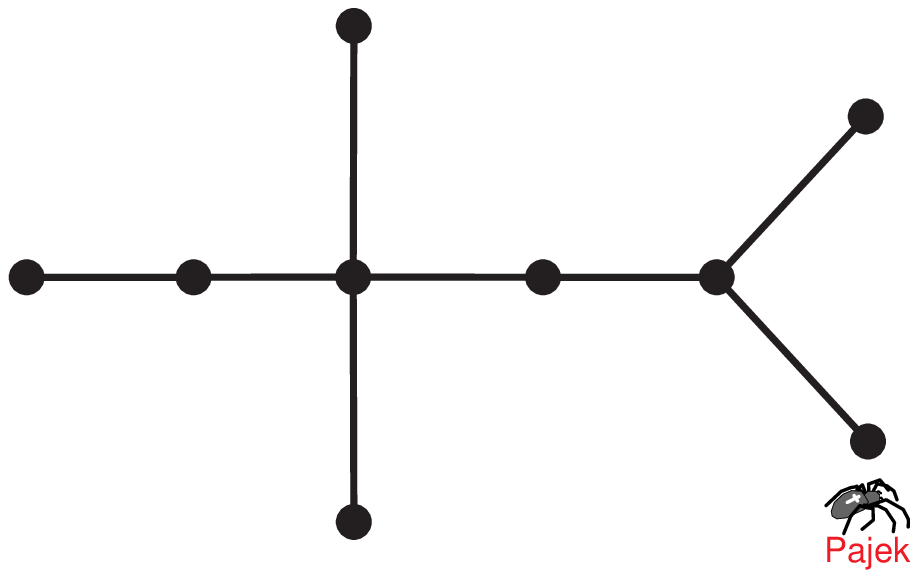}
\includegraphics[scale=.25]{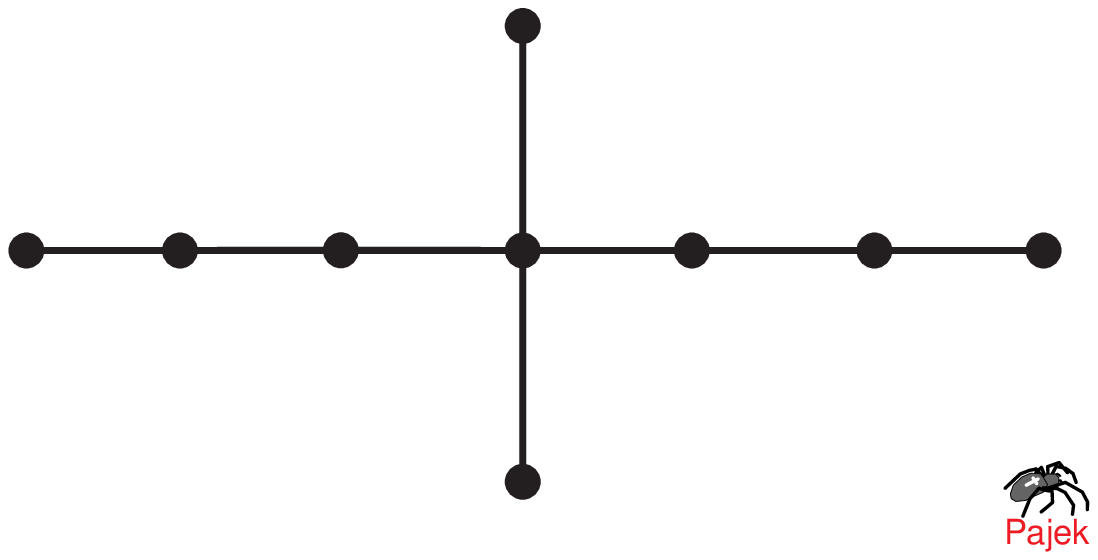}
\includegraphics[scale=.25]{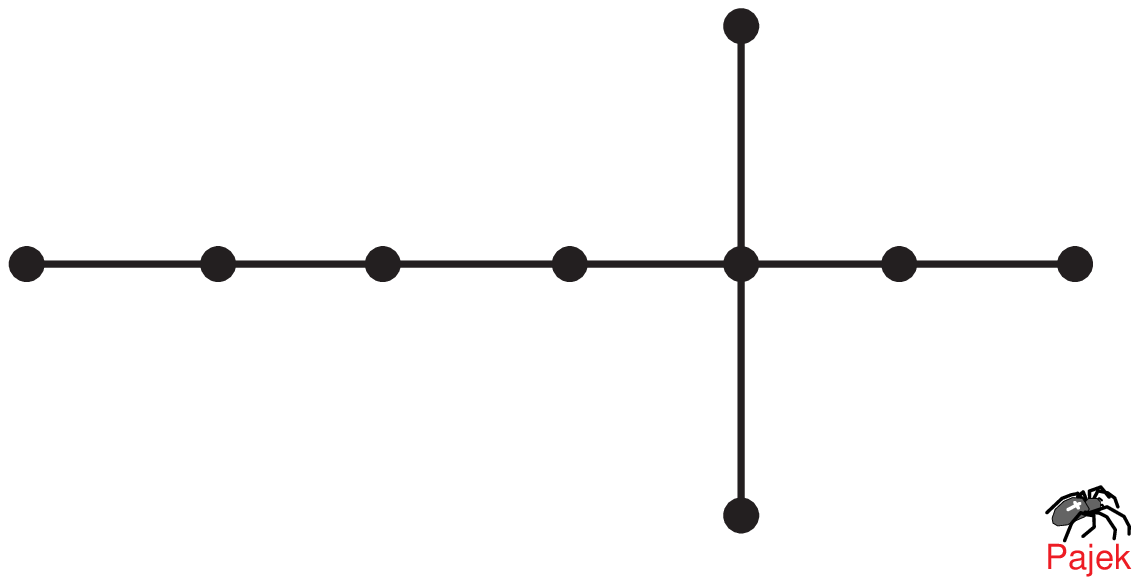}
\includegraphics[scale=.25]{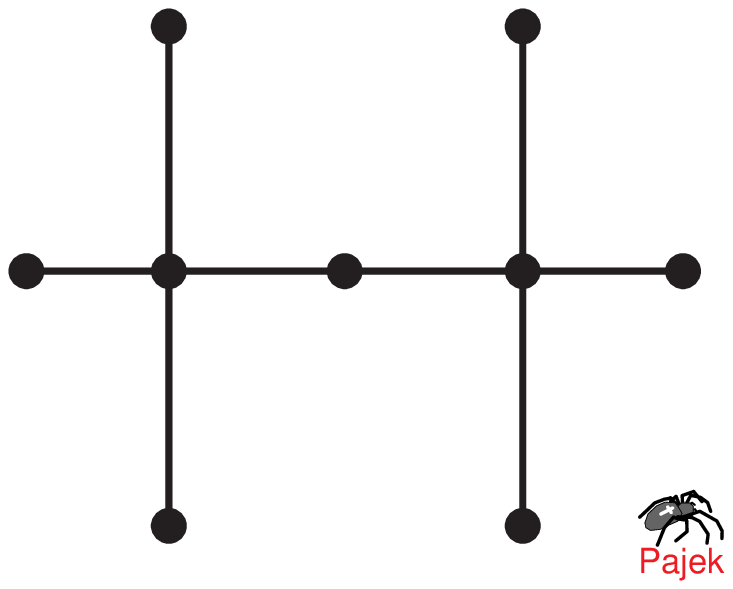}
\includegraphics[scale=.25]{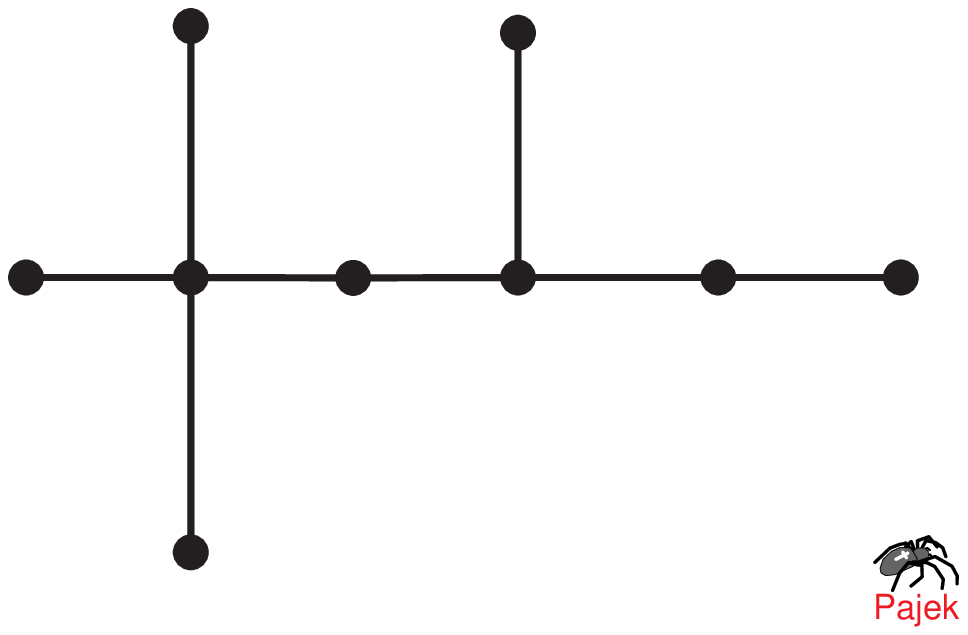}
\includegraphics[scale=.25]{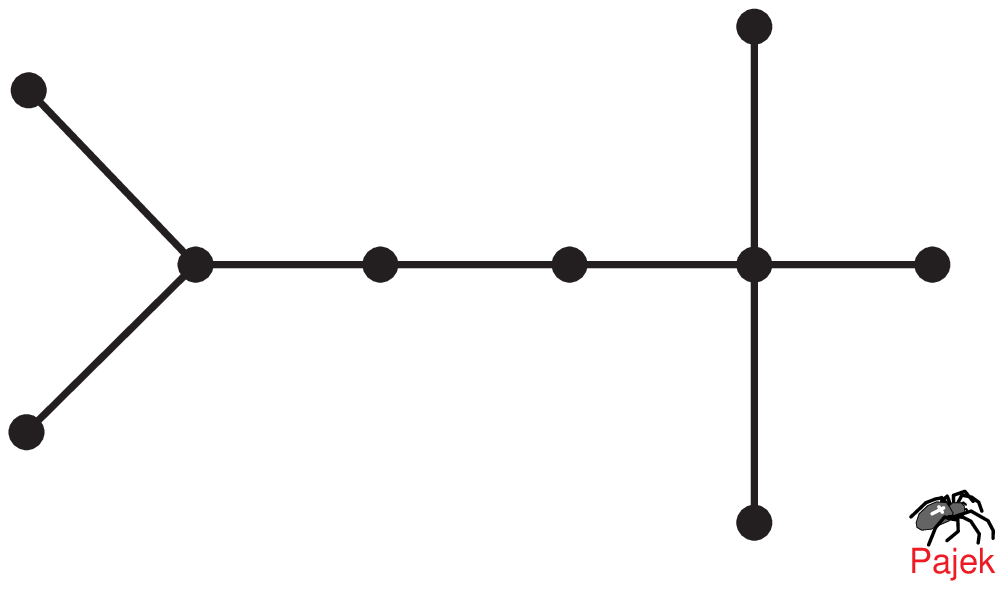}
\includegraphics[scale=.25]{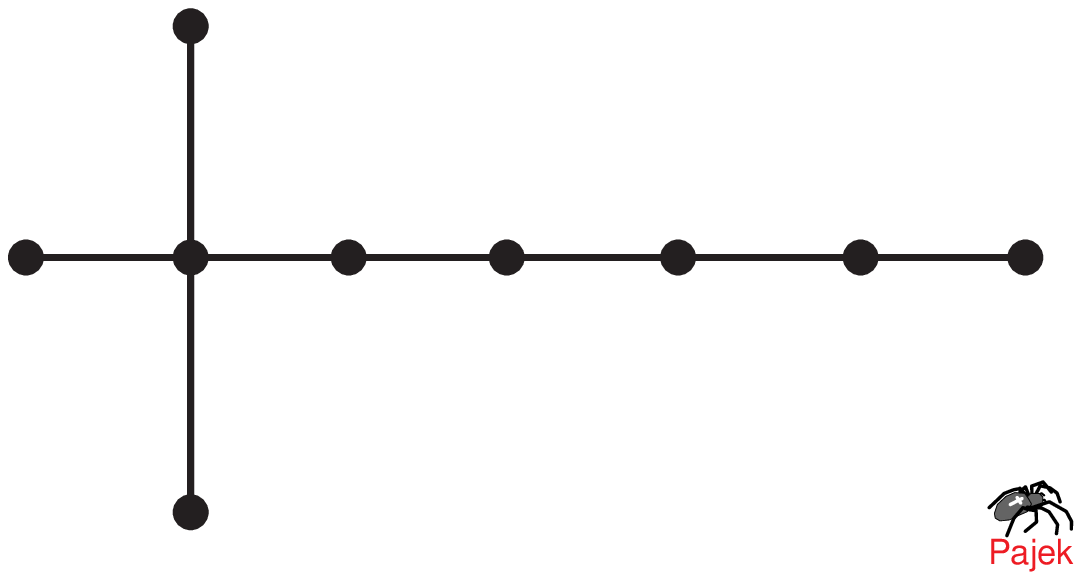}
\includegraphics[scale=.25]{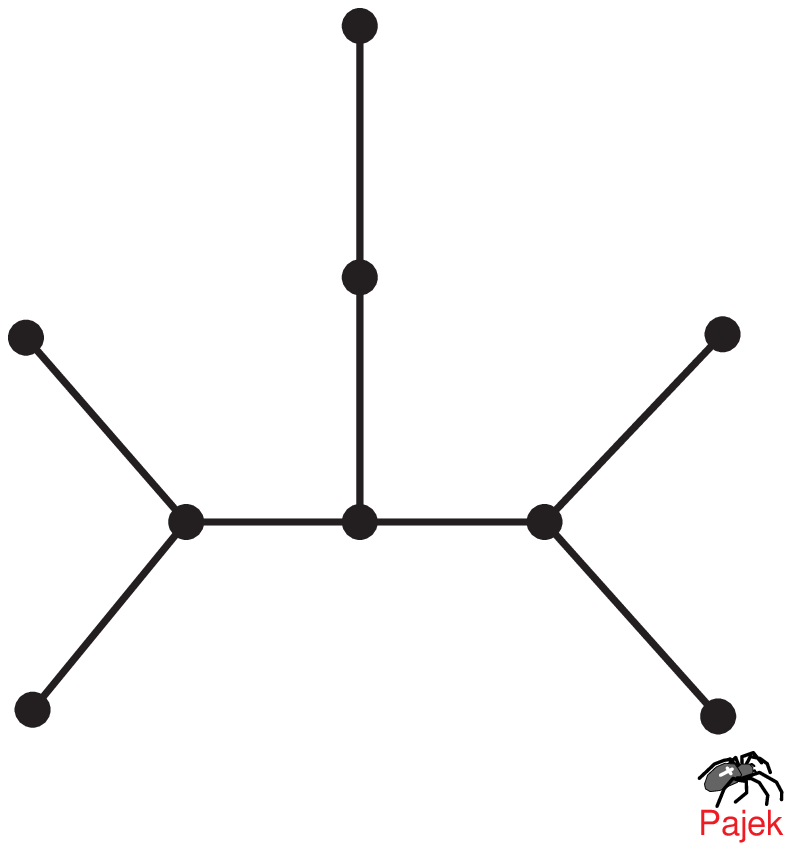}
\includegraphics[scale=.25]{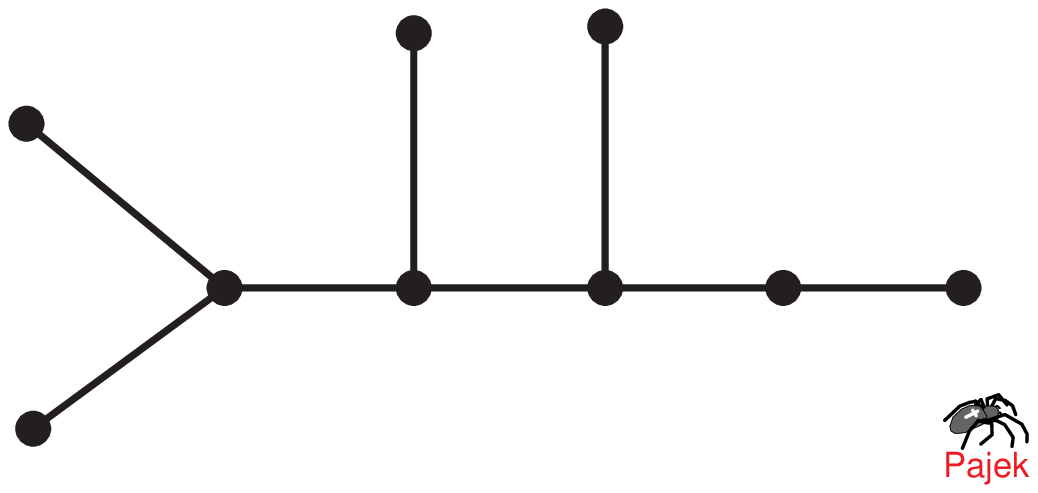}
\includegraphics[scale=.25]{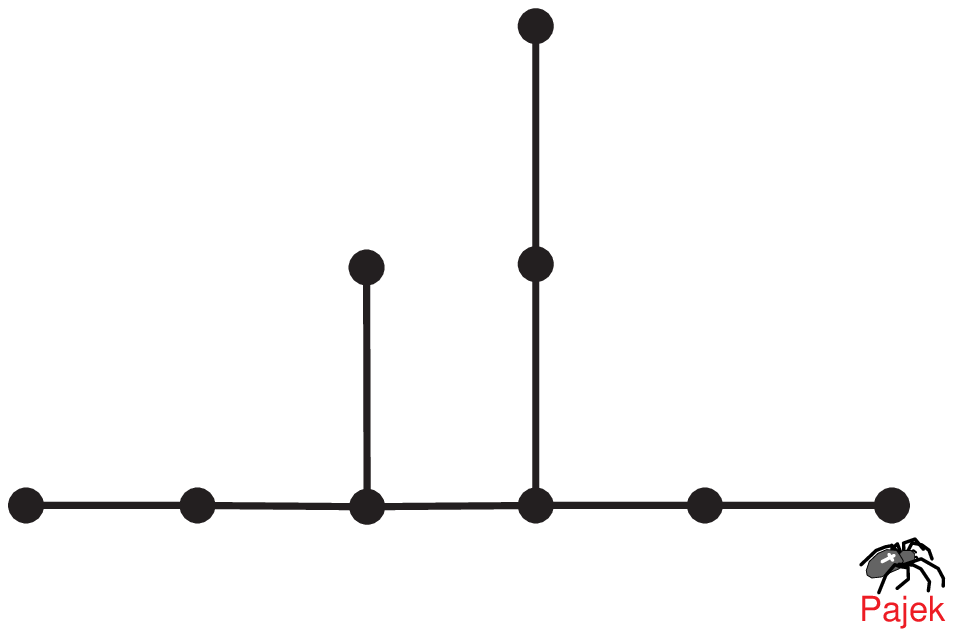}
\includegraphics[scale=.25]{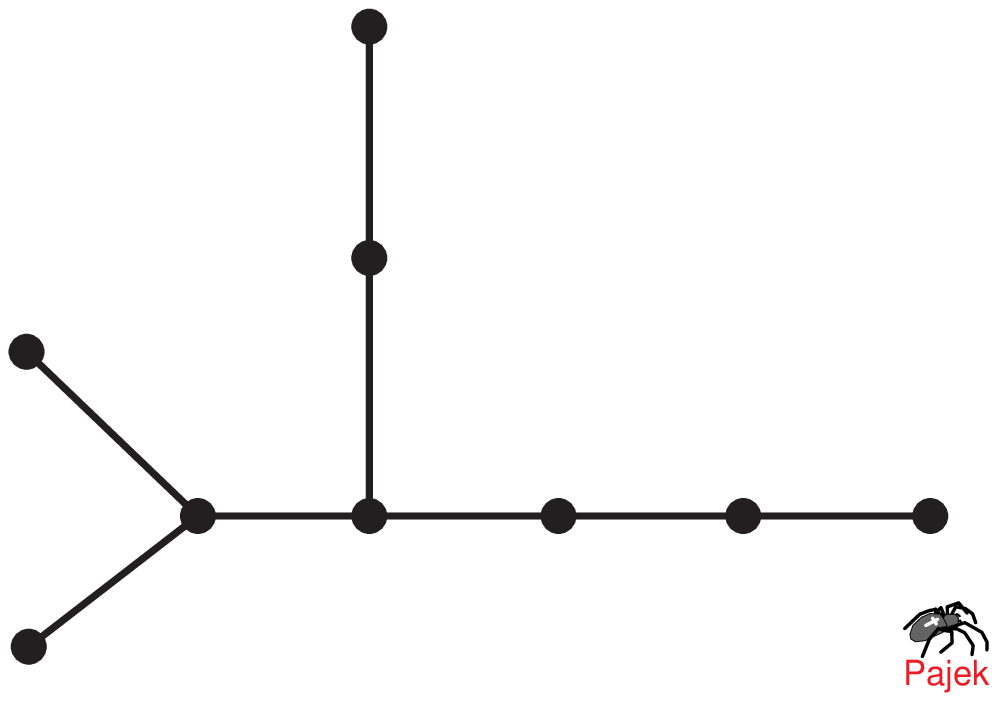}
\includegraphics[scale=.25]{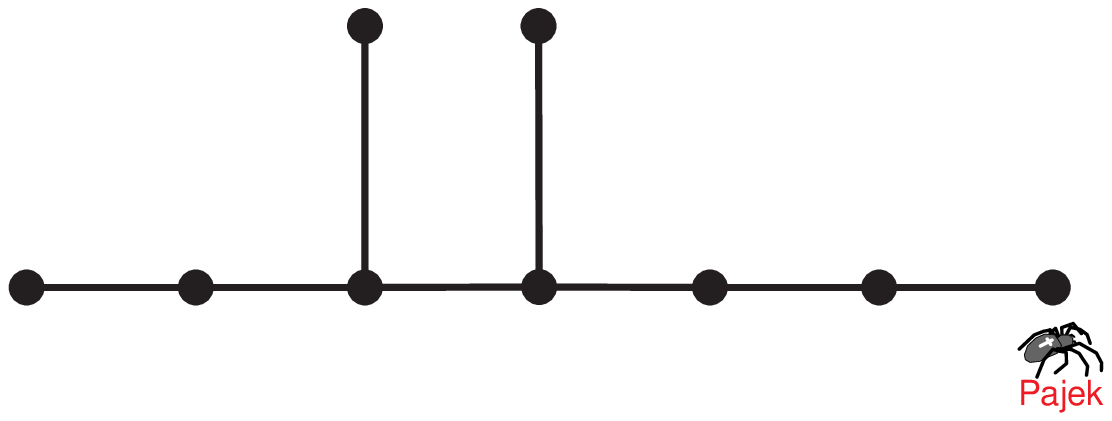}
\includegraphics[scale=.25]{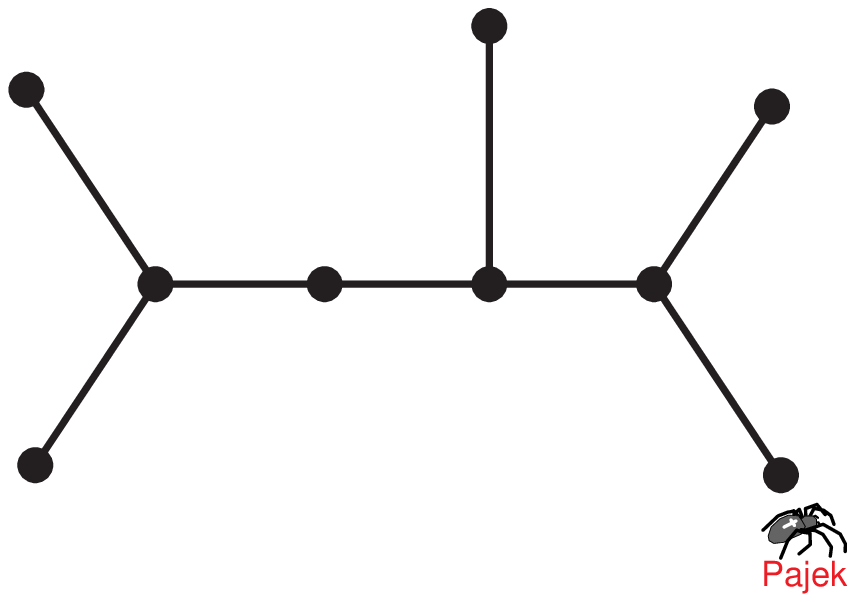}
\includegraphics[scale=.25]{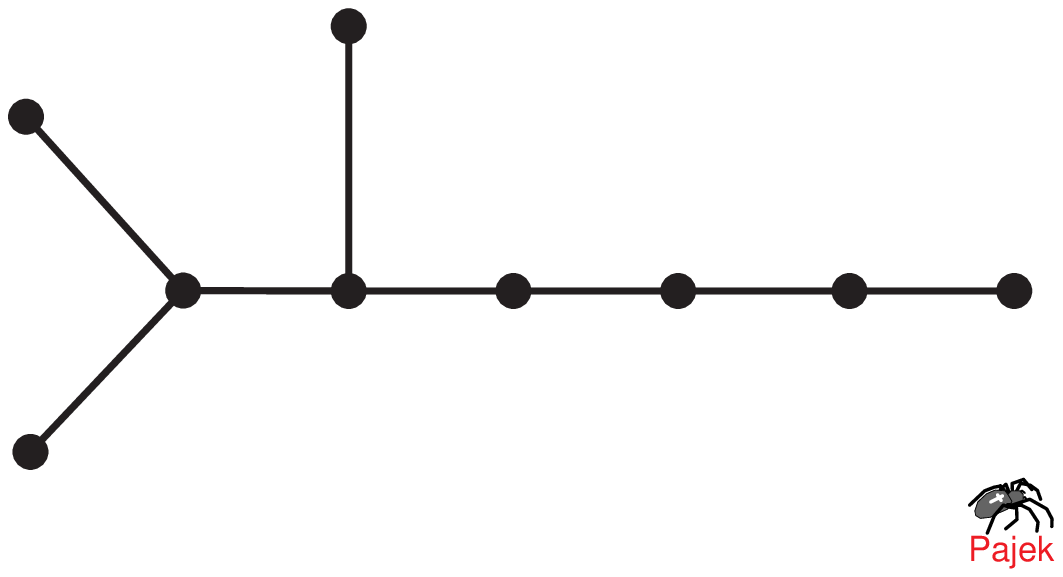}
\includegraphics[scale=.25]{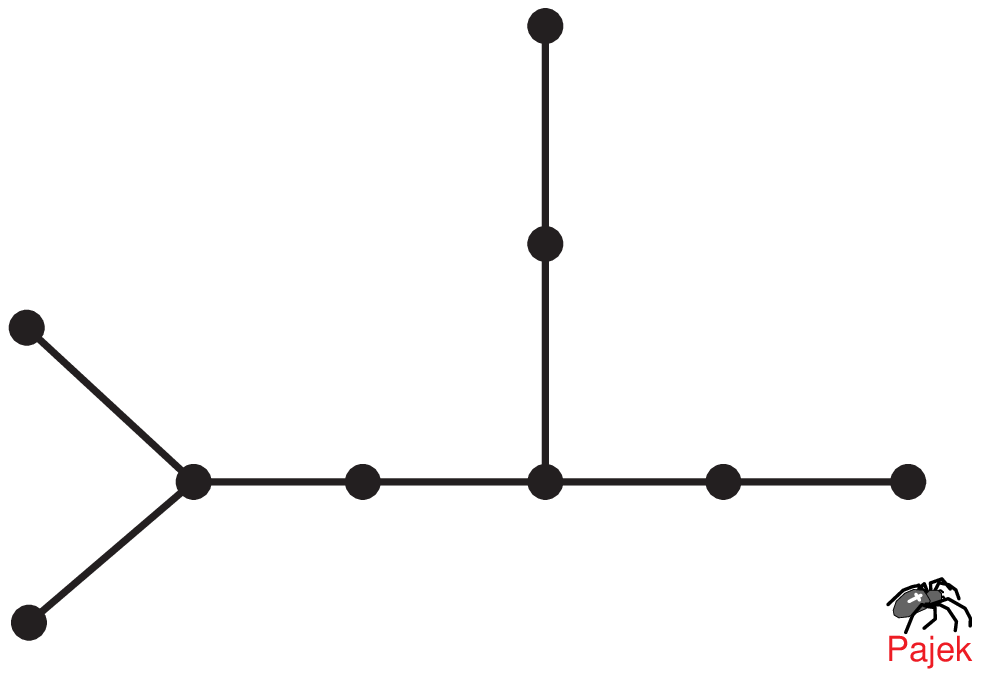}
\includegraphics[scale=.25]{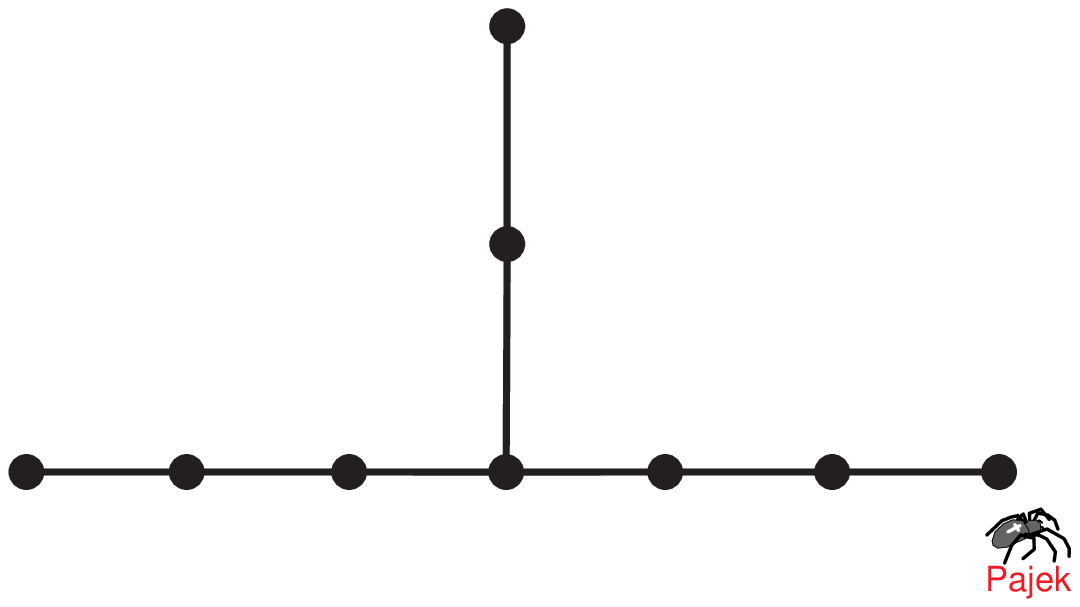}
\includegraphics[scale=.25]{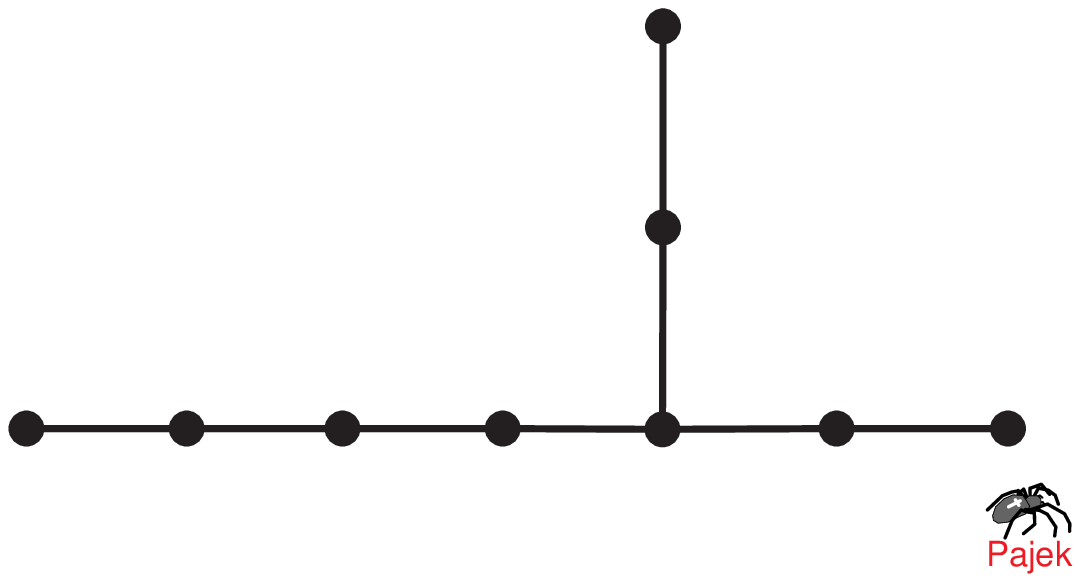}
\includegraphics[scale=.25]{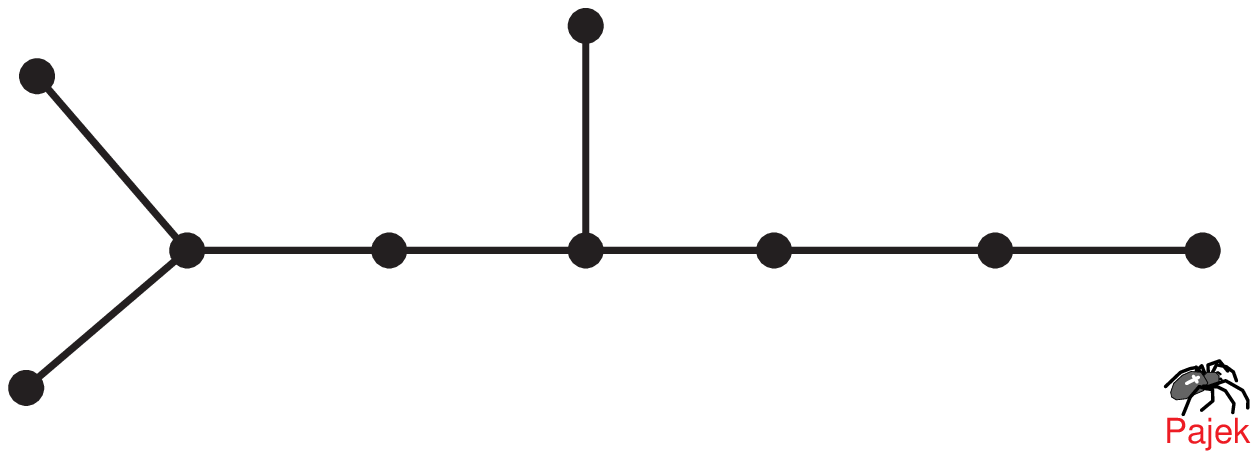}
\includegraphics[scale=.25]{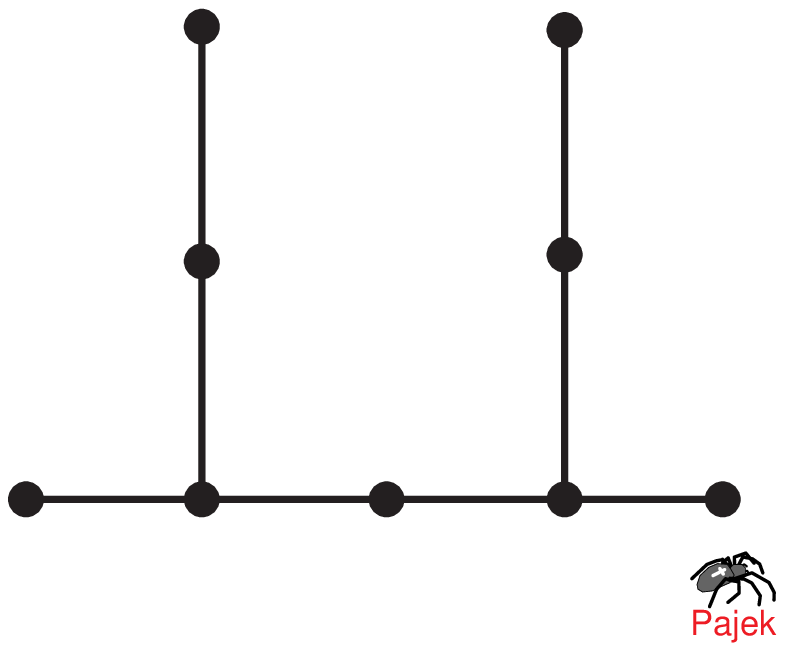}
\includegraphics[scale=.25]{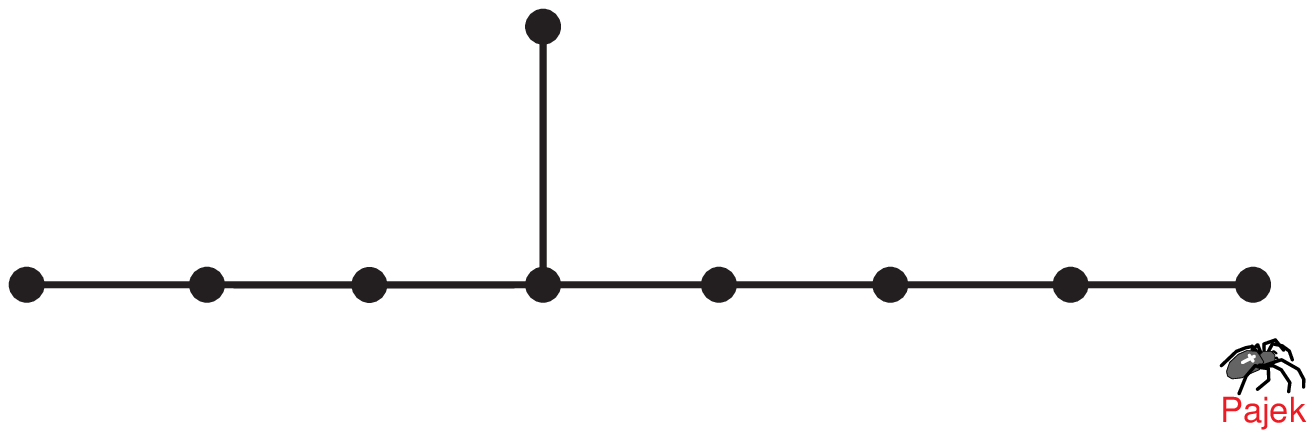}
\includegraphics[scale=.25]{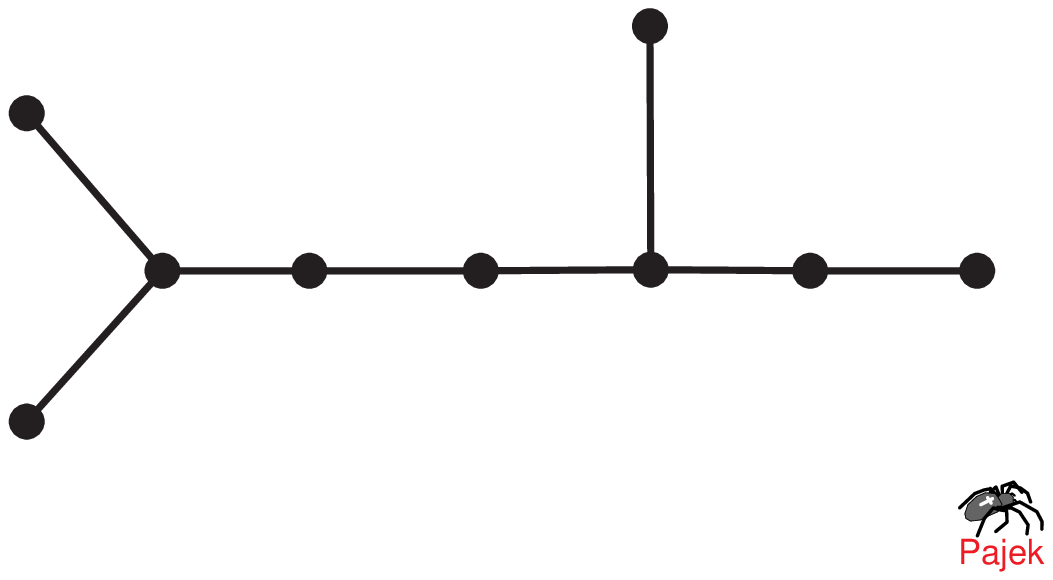}
\includegraphics[scale=.25]{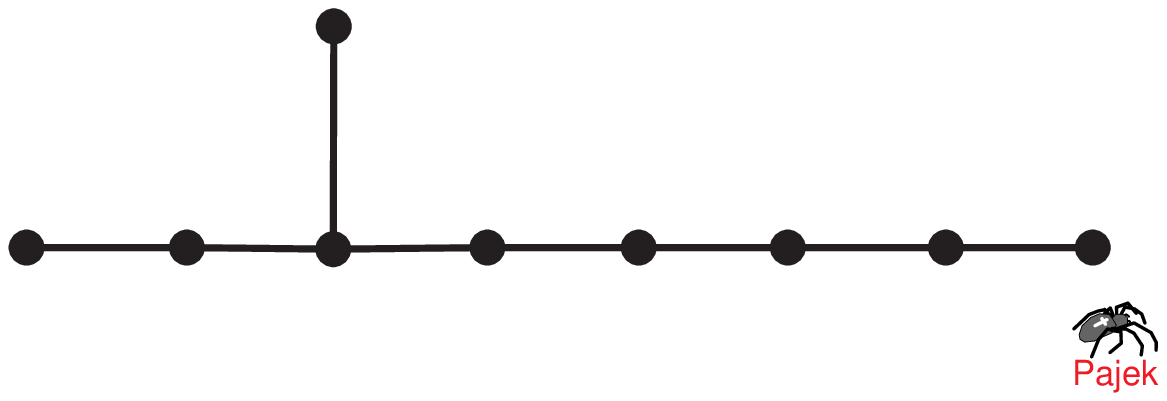}
\includegraphics[scale=.25]{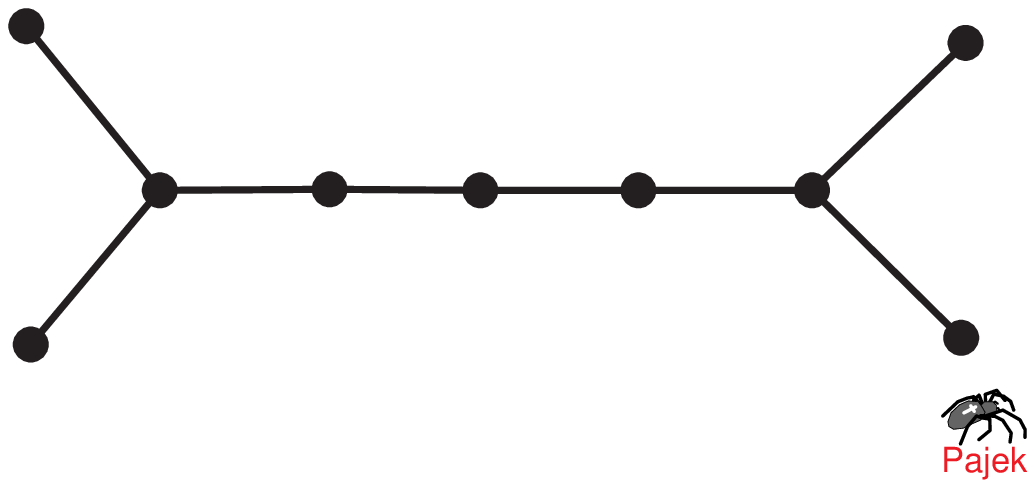}
\includegraphics[scale=.25]{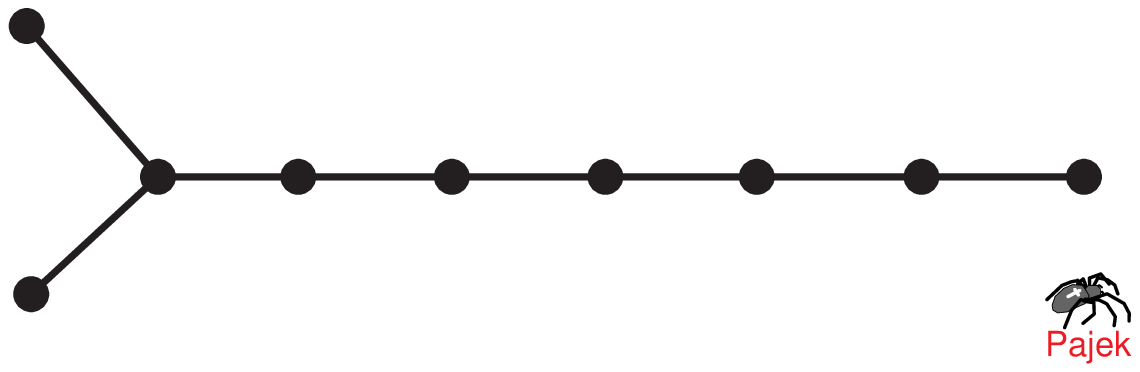}
\includegraphics[scale=.25]{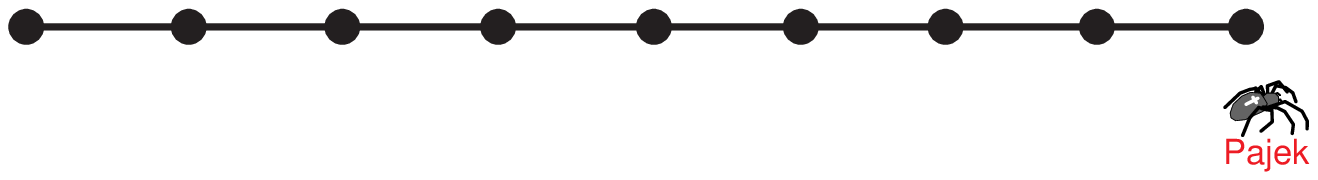}
\end{center}
\caption{\label{fig-trees9} All $T=47$ trees with $N=9$ nodes.}
\end{figure*}

The number of trees $T$ obtained with above algorithm with $k\le 5$ are given in Table \ref{tab}.
The results agree with the available number of trees given in Refs. \cite{tab8,net}.
For example, all $T=47$ trees of $N=9$ nodes are presented in Fig. \ref{fig-trees9}.

\begin{figure}
\begin{center}
\includegraphics[width=.9\textwidth]{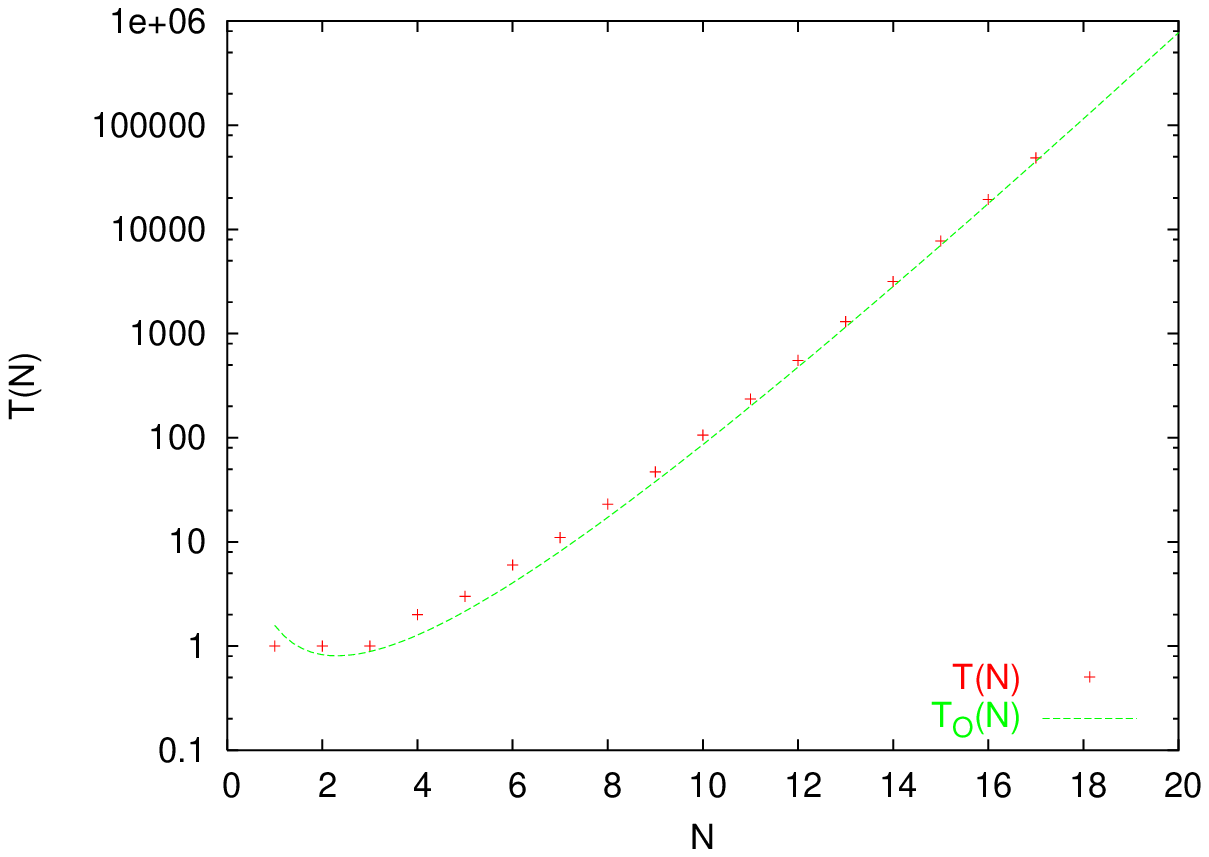}
\end{center}
\caption{\label{fig-otter} Number of trees $T$ as compared with the Otter's formula 
$T_{\text{O}}$.}
\end{figure}

For large enough $N$ the number of trees $T$ is asymptotically given as
\begin{equation}
\label{eq-otter}
T_{\text{O}}(N)=\beta \cdot \alpha^N \cdot N^{-5/2},
\end{equation}
where
$\alpha=2.9557652856\cdots$
and
$\beta=0.5349496061\cdots$
\cite{otter}.
The comparison of the results of the exact trees counting and predictions of Eq. 
\eqref{eq-otter} is shown in Table \ref{tab} and in Fig. \ref{fig-otter}.

In the terminology of Ref. \cite{szulc} the degree vector ${\bf v}$ is called {\em valence} 
vector.
The molecular topological index (MTI) is defined as\footnote{MTI was originally defined as a simple sum of elements of product ${\bf v}({\bf A}+{\bf D})$ and not as sum of absolute values of its elements.
As elements of ${\bf v}({\bf A}+{\bf D})$ are always positive our description is only more formally compact.}
\begin{equation}
\text{MTI}=\left\| {\bf v}\cdot({\bf A}+{\bf D}) \right\|,
\end{equation}
where
${\bf A}$ is a graph's adjacency matrix
and vector norm $||\cdots||$ is defined as sum of absolute value of vectors element
\[
\left\|{\bf c}\right\| = \left\|(c_1,c_2,\cdots,c_{N-1},c_N )\right\|\equiv\sum_{i=1}^N |c_i|.
\]
In adjacency matrix ${\bf A}$ element $a_{ij}$ gives number of edges between nodes $i$ and $j$.
For simple graphs --- where multiple edges are forbidden --- matrix ${\bf A}$ becomes binary.

\begin{figure}
\begin{center}
\begin{tabular}{r cc}
\hline \hline
MTI & & \\
230 & (a) \includegraphics[scale=0.3]{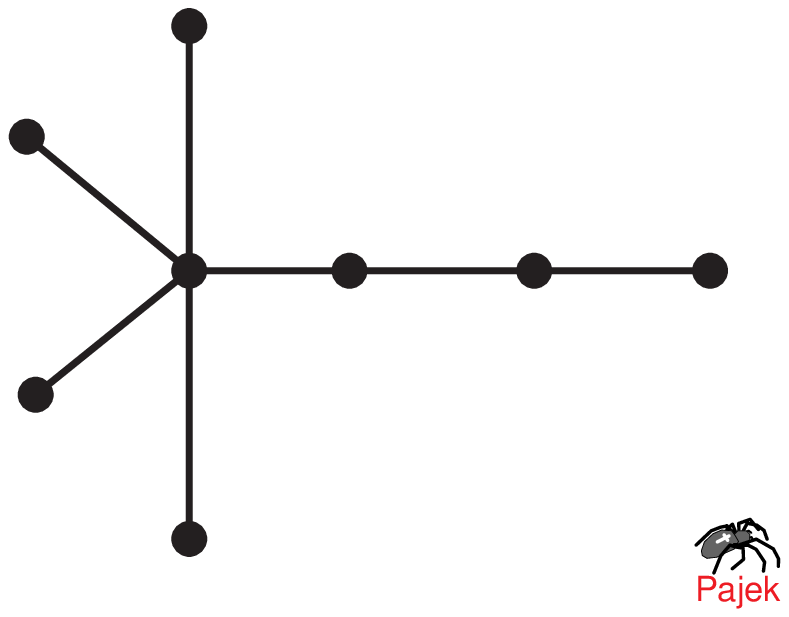}
    & (b) \includegraphics[scale=0.3]{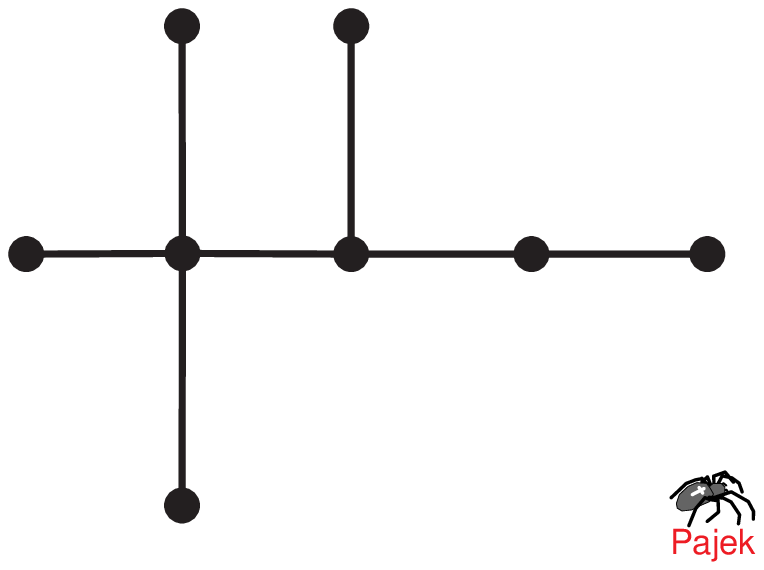} \\
242 & (c) \includegraphics[scale=0.3]{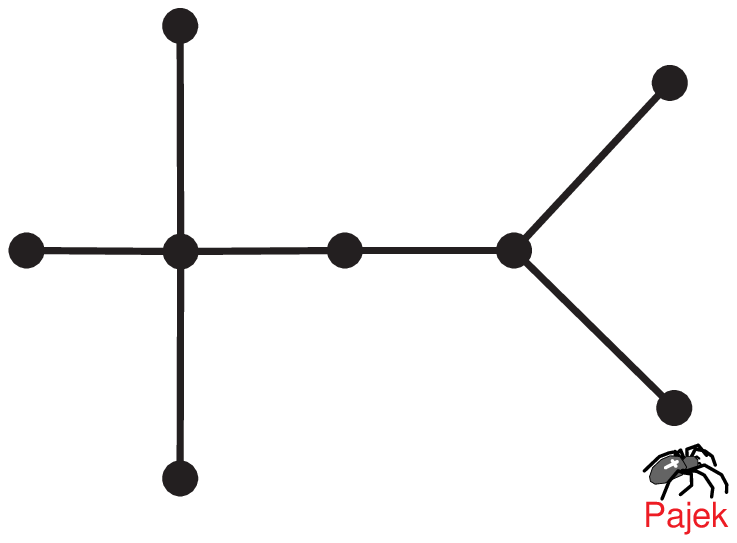}
    & (d) \includegraphics[scale=0.3]{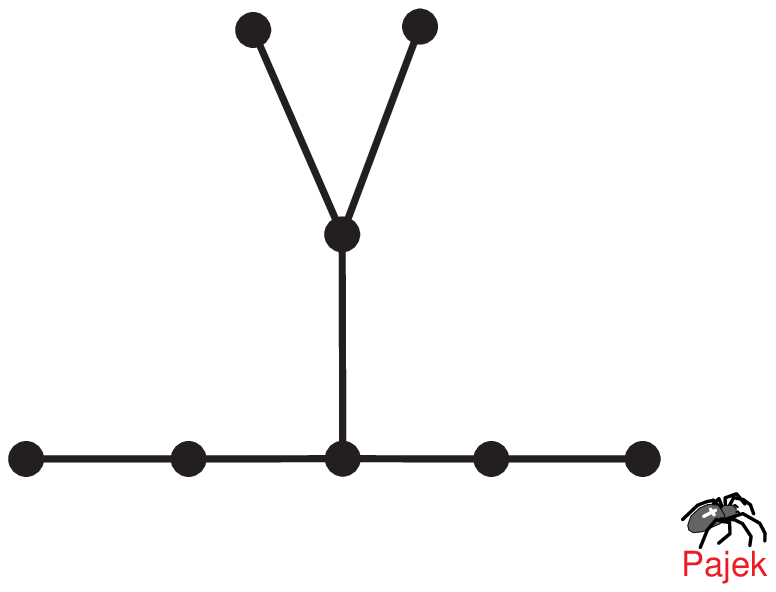} \\
260 & (e) \includegraphics[scale=0.3]{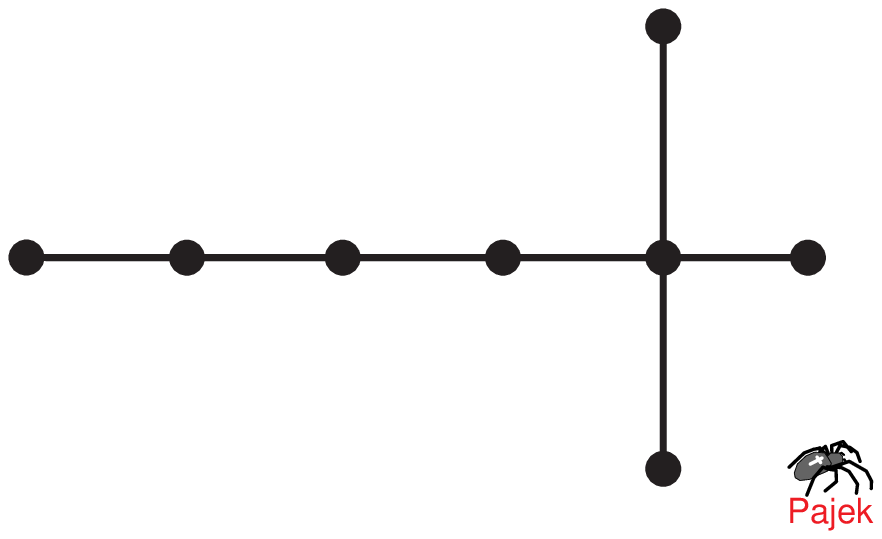} 
    & (f) \includegraphics[scale=0.3]{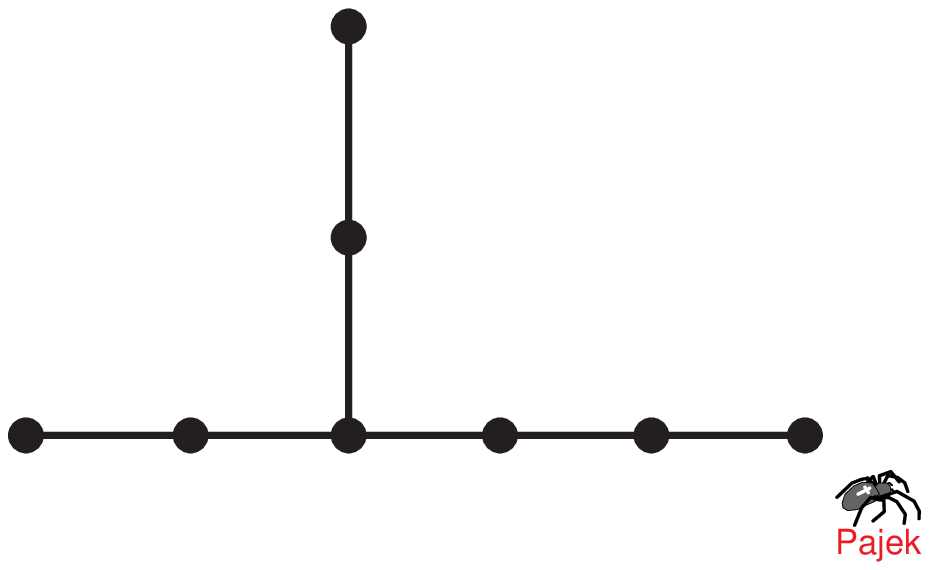} \\ 
\hline \hline
\end{tabular}
\end{center}
\caption{\label{fig-missing} Three pairs of different topologically trees with $N=8$ nodes with the same MTI.}
\end{figure}

MTI was believed to be single-number value which allow to differ between trees \cite{szulc}.
Here, however we can see that this method of counting fails for $N\ge 8$.  The obtained 
number of trees with Schultz method is $T_\text{S}(8)=20$ while true value is $T(8)=23$.
Three pairs of trees which have the same MTI but different $({\bf b},{\bf v})^8$ are shown in 
Fig. \ref{fig-missing}.

The purpose of introducing MTI was to differentiate chemical molecules.
When a carbon atom (with proper number of hydrogen atoms) is assigned to all nodes of trees shown 
in Fig. \ref{fig-missing}(c)-(f) they may represent semi-structural formulas of 
(c) 2,2,4-trimethylpentane, (d) 3-ethyl-2-methylpentane, (e) 2,2-dimethylhexane and 
(f) 3-ethylhexane \cite{chem}.
The MTI cannot differ between pairs (c,d) and (e,f) of these forms of $\text{C}_8\text{H}_{18}$.

Our results contain not only the number of trees, but the structure of all of them.
Binary files with distance matrices and the program for their conversion to input files for Pajek 
\cite{pajek} program are available from our web page \cite{www}.

\section{Discussion}
\label{sec-4}

Now we are going to prove that for large $N$, the range of any discriminative 
topological invariant with integer values should increase exponentially with $N$.
To each tree, a different value of the invariant must be assigned, 
if the invariant is discriminative. Then we get an exponentially 
increasing number of different integer values. The length of a
range on an axis, where these values can be placed, must increase 
also at least exponentially, what finishes the proof. We note that the
matrix character of the invariant does not change this result, as long
as the matrix size increases as $N^c$, where $c$ is a constant. In our case $c=1$,
because the matrix is $N\times 2$. We should add that this `range criterion' is
crucial in the asymptotic regime of large $N$. Up to now, the computational resources do not allow 
to penetrate this region.

Concluding, we have proposed a new topological invariant to discriminate unlabeled trees.
The matrix character of the invariant allows to believe, that the discriminating power of the 
invariant is much better, than scalar invariants proposed previously.

\ack
K.M. thanks Aleksandra Jung for her valuable help.
Calculations were carried out in ACK-CYFRONET-AGH.
The machine time on SGI 2800 is financed by the Polish Ministry of Science and 
Information Technology under Grant No. KBN/SGI2800/AGH/018/2003.


\end{document}